\documentstyle[preprint,aps,tighten]{revtex}
\begin{document}
\draft

\tolerance=5000
\def\pp{{\, \mid \hskip -1.5mm =}}
\def\cL{{\cal L}}
\def\be{\begin{equation}}
\def\ee{\end{equation}}
\def\bea{\begin{eqnarray}}
\def\eea{\end{eqnarray}}
\def\tr{{\rm tr}\, }
\def\nn{\nonumber \\}
\def\e{{\rm e}}
\def\D{{D \hskip -3mm /\,}}

\def\SEH{S_{\rm EH}}
\def\SGH{S_{\rm GH}}
\def\AdS5{{{\rm AdS}_5}}
\def\S4{{{\rm S}_4}}
\def\gfv{{g_{(5)}}}
\def\gfr{{g_{(4)}}}
\def\SC{{S_{\rm C}}}
\def\RH{{R_{\rm H}}}

\def\wlBox{\mbox{
\raisebox{0.1cm}{$\widetilde{\mbox{\raisebox{-0.1cm}\fbox{\ }}}$}}}
\def\htBox{\mbox{
\raisebox{0.1cm}{$\hat{\mbox{\raisebox{-0.1cm}{$\Box$}}}$}}}

\title{Cosmological and black hole brane-world Universes 
in higher derivative gravity }

\author{Shin'ichi Nojiri\thanks{Electronic address: 
snojiri@yukawa.kyoto-u.ac.jp, nojiri@cc.nda.ac.jp}}
\address{Department of Applied Physics,
National Defence Academy,
Hashirimizu Yokosuka 239-8686, JAPAN}

\author{Sergei D. Odintsov\thanks{On leave from Tomsk State 
Pedagogical University, 
634041 Tomsk, RUSSIA.
Electronic address: odintsov@ifug5.ugto.mx}}
\address{
Instituto de Fisica de la Universidad de Guanajuato,
Lomas del Bosque 103, Apdo.\ Postal E-143, 
37150 Leon, Gto., MEXICO}

\author{Sachiko Ogushi\thanks{JSPS fellow, 
Electronic address: ogushi@yukawa.kyoto-u.ac.jp}}
\address{Yukawa Institute for Theoretical Physics, 
Kyoto University, Kyoto 606-8502, JAPAN}

\maketitle
\begin{abstract}
General model of multidimensional $R^2$-gravity including Riemann 
tensor square term (non-zero $c$ case) is considered.
The number of brane-worlds in such model is constructed 
(mainly in five dimensions) and their properties are discussed. 
Thermodynamics of S-AdS BH (with boundary) is presented when 
perturbation on $c$ is used. The entropy, free energy 
and energy are calculated. For non-zero $c$ the entropy (energy) 
is not proportional to the area (mass). The equation of motion 
of brane in BH background is presented as FRW equation. Using 
dual CFT description it is shown that dual field theory is not 
conformal one when $c$ is not zero. In this case the holographic 
entropy does not coincide with BH entropy (they coincide for 
Einstein gravity or $c=0$ HD gravity where AdS/CFT 
description is well applied).

Asymmetrically warped background (analog of charged AdS BH) 
where Lorentz invariance violation occurs is found. The 
cosmological 4d dS brane connecting two dS bulk spaces is 
formulated in terms of parameters of $R^2$-gravity. Within 
proposed dS/CFT correspondence the holographic conformal 
anomaly from five-dimensional higher derivative 
gravity in de Sitter background is evaluated.

\end{abstract}


\section{Introduction}

It is quite an old idea going back to Kaluza and Klein that theory 
of fundamental interactions is some manifestation of the presence of 
extra dimensions. The investigation of higher dimensional theories 
has appeared recently in the form of brane-world 
physics \cite{RS,rsrelated}.
According to this (simplified) picture the fundamental dynamics 
occurs in $d+1$ dimensional bulk manifold with $d$ dimensional 
boundary (brane).
The evolution of observable universe corresponds to brane.

Brane-world description \cite{RS} looks really attractive due to the 
following reasons. First of all, in most models the gravity on 
the brane is localized \cite{RS}. In other words, despite the fact 
that gravity is 
multidimensional one, the Newton law is recovered on the brane. 
It happens even with large (infinite) extra dimensions. 
Moreover, brane matter may be localized as well. 
Second, this approach provides very natural explanation of hierarchy 
between the gravitational and electroweak mass scales. It opens 
the window for construction of new classes of unified models from
higher dimensional gravity.
Third, standard cosmology is modified by the presence of extra 
dimensions. 
Moreover, standard Friedmann-Robertson-Walker (FRW) cosmology is 
recovered at low energies. On the same time the role of brane-world 
effects to
very early universe may be quite significant. In particulary, higher 
dimensions effects may give the origin of dark matter. There is much 
activity in the study of brane-world cosmology, see 
refs.\cite{cosmology,cosmology2,strings}.
Fourth, brane-world physics very naturally appears in frames of 
string and M-theory. For example, some brane-world cosmological 
scenarios (like Brane New World \cite{NOZ,HHR0,related}) are 
very much connected with AdS/CFT correspondence \cite{AdS}. 
Moreover, such approach is really useful to test the holography 
predictions. In a sense, the very much related elements of
new physics (holographic principle, AdS/CFT correspondence, 
brane-worlds) appeared almost simultaneously.

The priority area is currently the search of realistic brane 
universes within different theories. The most attention is 
paid so far to Einstein gravity (with matter). Nevertheless, 
there is interest in the study of brane universes in higher 
derivative (HD) gravity \cite{HDt}. It is not strange as 
usually at low energies HD gravity approaches to Einstein one, 
i.e. the difference between these theories is typical only at 
high energies where brane universe occurs. It is known that 
four dimensional HD gravity 
has better ultraviolet behaviour than Einstein one.
HD gravitational terms appear in low-energy effective action 
of string theory as well as in holographic renormalization 
group \cite{BVV}.
Moreover, some variants of HD gravity may possess the interesting 
additional symmetry (Weyl gravity). Gauss-Bonnet (GB) gravity 
which is another variant of HD gravity is typically associated with 
superstring\cite{GBaction}. It has very attractive ultraviolet 
behaviour of propagator for number of backgrounds in $d$ dimensions.

The theory we are going to investigate in this work in connection 
with brane-world approach is multidimensional HD 
(or $R^2$-) gravity. The construction of different warped 
spacetimes (brane-worlds) in HD gravity will be given 
and a number of related questions (thermodynamics of AdS black 
holes (BH), FRW brane dynamics, de Sitter (dS) brane, etc) 
will be discussed.
The remark may be in order. In four dimensions where a lot of works 
devoted to HD gravity exists (for review, see \cite{BOS}) one 
only needs scalar curvature square and Ricci tensor square terms. 
Riemann tensor square term may be always discarded using GB 
topological invariant. In $d$ dimensions, Riemann tensor square 
term should be kept and as we will see it leads to some new 
features in the description of AdS BH thermodynamics.

The paper is organized as follows. In the next section the 
equations of motion for general HD gravity are reviewed in the 
case of FRW background. 
Section three is devoted to the construction of boundary terms in 
$d$ dimensional HD gravity. The role of these terms is to make the 
variational procedure to be well-defined and to render the classical 
action to be finite for AdS background. In a sense, such 
construction is the generalization of Gibbons-Hawking boundary 
term \cite{GB} in Einstein gravity. 

In the section four we study thermodynamics of 
five dimensional AdS BH (with brane) in 
general model of HD gravity. The entropy, free energy and 
energy of AdS BH are obtained and compared with the same ones 
calculated by another method. It is demonstrated that for AdS BH 
with curved brane the entropy is not proportional to horizon 
area only if Riemann tensor square term 
presents in the action. This new, normally non-observed feature of 
the entropy is presumbly caused by above HD correction. 
In the section five the derivation of brane equation (FRW 
equation) when brane sits inside AdS BH is done. The dual 
AdS/CFT description is used. It is demonstrated that presence 
of Riemann tensor square term breaks the 
conformal symmetry of dual field theory. It results in the fact that 
there appears the non-trivial difference between holographic (Hubble)  
QFT entropy (when brane crosses the horizon) and black hole entropy. 
In the absence of Riemann tensor square term these entropies 
coincide for HD as well as for Einstein gravity \cite{EV}. 
In the section six we re-write the results of two previous 
sections about AdS BH thermodynamics and brane 
equation for Gauss-Bonnett gravity.

Section seven is devoted to construction of asymmetrically 
warped spacetime in general HD gravity. Riemann tensor square 
term produces such background in the way similar to 
Einstein-Maxwell gravity, however, in our case the Maxwell 
charge contribution is not necessary. The apparent violation of 
Lorentz invariance for such background is briefly mentioned. 

In the section eight we discuss dS brane solutions in bulk 
de Sitter space. The explicit construction of such brane-worlds 
motivated by dS/CFT correspondence is presented. In section nine, 
assuming dS/CFT correspondence the calculation of central 
charge (holographic conformal anomaly) from five dimensional 
HD gravity on dS space is made. In the absence of string 
framework for dS/CFT correspondence, (somehow speculative)
example of specific five dimensional de Sitter HD gravity model 
which has dual $Sp(N)$ ${\cal N}=2$ super Yang-Mills theory 
description up to the next-to-leading terms in $1/N$ expansion 
is presented. Summary of results and outlook are  given in the last 
section. Appendix is devoted to description of AdS BH 
thermodynamics for 
HD gravity without Riemann tensor square term.

\section{FRW dynamics from multidimensional $R^2$-gravity}

HD gravity attracts the attention of researchers due to
different reasons. Clearly, this is very natural generalization 
of Einstein gravity. Moreover, HD terms appear in
low-energy effective action of superstring theory.
On the quantum level, HD gravity seems to be renormalizable in 
four dimensions (for a review, see \cite{BOS}) forgetting 
the unitarity problem (for recent discussion of life with 
ghosts in HD theories, see\cite{unit}).

HD gravity appears also very naturally in brane-world physics 
and AdS/CFT set-up. Indeed, imagine one  substitutes the 
classical solution into the action of the supergravity on 
AdS background. The action diverges due to the infinite volume 
of  AdS. In order that AdS$_{d+1}$/CFT$_d$ correspondence \cite{AdS} 
was well-defined, we need to add the surface term to cancell 
the divergence. 
The leading divergence has the form of the cosmological term 
in $d$ dimensional spacetime and the next-to-leading term is 
Einstein-Hilbert action (scalar curvature term). 
The next-to-next-to-leading terms contain the square of the curvatures, 
which correspond to the trace anomaly when $d=4$. 
In the context of the Brane New World \cite{NOZ,HHR0}, 
the gravity localized on the brane in the Randall-Sundrum 
model \cite{RS} corresponds to the remnant after cancelling 
the leading divergent cosmological-like term. Therefore 
the gravity on the brane always contains the 
$R^2$-graviy as the correction to the Einstein gravity. 

Let us review FRW-dynamics in general multidimensional HD gravity.
The general action of $d$ dimensional $R^2$-gravity with 
cosmological constant and matter is given by:
\be
\label{vi}
S=\int d^d x \sqrt{-g}\left\{a R^2 + b R_{\mu\nu} R^{\mu\nu}
+ c  R_{\mu\nu\xi\sigma} R^{\mu\nu\xi\sigma}
+ {1 \over \kappa^2} R - \Lambda 
+ L_{\rm matter}\right\}\ .
\ee
Here $L_{\rm matter}$ is the Lagrangian density for the matter 
fields. 
By the variation over the metric tensor $g_{\mu\nu}$, we obtain 
the following equation
\bea
\label{R1}
0&=&{1 \over 2}g^{\mu\nu}\left(a R^2 + b R_{\mu\nu} R^{\mu\nu}
+ c R_{\mu\nu\xi\sigma} R^{\mu\nu\xi\sigma}
+ {1 \over \kappa^2} R - \Lambda \right) \nn
&& + a\left(-2RR^{\mu\nu} + \nabla^\mu\nabla^\nu R 
+\nabla^\nu\nabla^\mu R 
 - 2g^{\mu\nu} \nabla_\rho\nabla^\rho R \right) \nn
&& + b\left( -2 R^\mu_{\ \rho} R^{\nu\rho} 
+ \nabla_\rho\nabla^\mu R^{\rho\nu}
+ \nabla_\rho\nabla^\nu R^{\rho\mu} - \Box R^{\mu\nu} 
 - g^{\mu\nu}\nabla^\rho\nabla^\sigma R_{\rho\sigma} \right) \nn
&& + c\left(-2R^{\mu\rho\sigma\tau}R^\nu_{\ \rho\sigma\tau}
 - 2 \nabla_\rho\nabla_\sigma R^{\mu\rho\nu\sigma}
 - 2 \nabla_\rho\nabla_\sigma R^{\nu\rho\mu\sigma}\right) \nn
&& - { 1 \over \kappa^2}R^{\mu\nu} - T_{\rm matter}^{\mu\nu} \ .
\eea
Here $T_{\rm matter}^{\mu\nu}$ is the energy-momentum tensor of 
the matter fields. 
As we are  interested in the cosmological problem, 
we choose the spacetime metric in the following form: 
\be
\label{R2}
ds^2 = - dt^2 + l^2 \e^{2A(t)}\tilde g_{ij}(x^k)dx^i dx^j\ .
\ee
Here $\tilde g_{ij}$ is the metric of the $d-1$ dimensional 
Euclidean Einstein manifold defined by
\be
\label{R3}
\tilde R_{ij}=k\tilde g_{ij}\ .
\ee
Here $k$ is a constant and $\tilde R_{ij}$ is the Ricci 
tensor defined by $\tilde g_{ij}$. In the following, we denote 
the quantities given by $\tilde g_{ij}$  using $\tilde{\ }$. 
The natural assumption is
\be
\label{R4}
\tilde R_{ijkl}={k \over d-1}\left(\tilde g_{ik}
\tilde g_{jl}  - \tilde g_{il}\tilde g_{jk} \right)\ .
\ee
Taking the metric as in (\ref{R2}), one gets the following 
expressions for the connection and the curvatures:
\bea
\label{R5}
&& \Gamma^t_{tt}=\Gamma^t_{ti}=\Gamma^t_{it}
=\Gamma^i_{tt}=0\ ,\quad 
\Gamma^t_{ij}=l^2\e^{2A}A_{,t}\tilde g_{ij}\ ,\nn
&& \Gamma^i_{jt}=\Gamma^i_{tj}=\delta^i_{\ j}A_{,t}\ ,
\quad \Gamma^i_{jk}=\tilde\Gamma^i_{jk}\ , \\
\label{R6}
&& R_{titj}=-R_{tijt}=-R_{ittj}=R_{itjt}
=-\e^{2A}\left(A_{,tt} + A_{,t}^2 \right)\tilde g_{ij}\ ,\nn
&& R_{ijkl}=\left({k \over d-2}l^2 \e^{2A} 
+ l^4 \e^{4A}A_{,t}^2\right)\left(\tilde g_{ik} 
\tilde g_{jl}  - \tilde g_{il} \tilde g_{jk} \right)\ ,\nn
&& \mbox{other components of Riemann tensor}=0\ , \\
\label{R7}
&& 
R_{tt}= - (d-1)\left(A_{,tt} + A_{,t}^2 \right) \ ,
\quad R_{ti}=R_{it}=0\ ,\nn
&& R_{ij}=\left\{k + l^2 \e^{2A}\left(A_{,tt} + (d-1)A_{,t}^2
\right)\right\}\tilde g_{ij}\ ,\\
\label{R8}
&& R = (d-1)kl^{-2}\e^{-2A} + 2(d-1)A_{,tt} + d(d-1)A_{,t}^2\ .
\eea
Here the following conventions of curvatures are used 
\bea
\label{curv}
R&=&g^{\mu\nu}R_{\mu\nu}\ , \nn
R_{\mu\nu}&=& {R^\lambda}_{\mu\lambda\nu}\ , \nn
{R^\lambda}_{\mu\rho\sigma}&=& -\Gamma^\lambda_{\mu\rho,\nu}
+ \Gamma^\lambda_{\mu\nu,\rho}
 - \Gamma^\eta_{\mu\rho}\Gamma^\lambda_{\nu\eta}
+ \Gamma^\eta_{\mu\nu}\Gamma^\lambda_{\rho\eta}\ , \nn
\Gamma^\eta_{\mu\lambda}&=&{1 \over 2}g^{\eta\nu}\left(
g_{\mu\nu,\lambda} + g_{\lambda\nu,\mu} - g_{\mu\lambda,\nu} 
\right)\ .
\eea
When $a=b=c=0$ the $(\mu,\nu)=(t,t)$ component in (\ref{R1}) 
corresponds to the first FRW equation and $(i,j)$ component to the 
second one. 

The $(t,t)$ component in (\ref{R1}) gives
\bea
\label{R8b}
0&=&-{(d-1)k \over 2\kappa^2 r^2} - {(d-1)(d-2) \over 2\kappa^2}
H^2 - \rho \nn
&& + \left( 2(d-1)^2 a+ {d(d-1) \over 2} b
+ 2(d-1) c\right) H_{,t}^2 \nn
&& + \left( - 4(d-1)^2 a -d(d-1)b - 4(d-1) \right) H H_{,tt} \nn
&& + \left(-4(d-1)^3 a -d(d-1)^2b -4 (d-1)^2\right) H^2 H_{,t} \nn
&& + \left( - {d(d-1)^2(d-4) \over 2} a - {(d-1)^2(d-4) \over 2} b 
 -(d-1)(d-4) c \right) H^4 \nn
&& + {k \over r^2} \left( - (d-6)(d-1)^2 a + 2(d-1) b 
 + 2(d-1) c \right) H^2 \nn
&& + \left( - {(d-1)^2 \over 2} a - {d-1 \over 2} b 
 - {d-1 \over d-2} c\right) {k^2 \over r^4}\ ,
\eea
which gives a generalization of the first FRW equation for 
the presence of HD terms. Here 
\be
\label{R9}
r=l\e^A\ ,\quad H=A_{,t}\ .
\ee
and $\rho$ is the energy density. 
On the other hand, $(\mu,\nu)=(i,j)$ components give
\bea
\label{R9b}
0&=&{1 \over \kappa^2}\left\{{(d-3)k \over 2r^4} 
 + {(d-2)H_{,t} \over r^2} 
 + {(d-1)(d-2) \over 2}H^2 \right\} - p \nn
&& + \left(4(d-1) a + db + 4c\right) H_{,ttt} \nn
&& + \left(8(d-1)^2 a + 2d(d-1) b + 8(d-1) c \right) H H_{,tt} \nn
&& + \left( 6(d-1)^2 a + {3d(d-1) \over 2} b 
+ 6(d-1)c \right) H_{,t}^2 \nn
&& + \left( (6d^2 - 16 d + 4)(d-1)a + (d^2 + d -8) (d-1) b 
\right. \nn
&& \left. + (4d^2 - 4d -12)c \right) H^2H_{,t} \nn
&& + \left( {d(d-1)^2(d-4) \over 2}a + {(d-4)(d-1)^2 \over 2}b 
+ (d-1)(d-4)c\right)  H^4 \nn
&& + {k \over r^2}\left\{\left(2(d-1)(d-6) a - 4b
 -4c\right)H_{,t} \right. \nn
&& \left. + \left( (d-1)(d^2 - 9d + 18) a
 -2(d-3)b -2(d-3) c \right) H^2 \right\} \nn
&& + {k^2 \over r^4}\left({(d-1)^2 \over 2}a + {d-1 \over 2}b 
+ {d-1 \over d-2}c\right)\ .
\eea
Here $p$ is the pressure. 
When $a=b=c=0$, Eqs.(\ref{R8b}) and (\ref{R9b}) reduce into the usual 
FRW equations 
\bea
\label{R8c}
0&=&-{(d-1)k \over 2\kappa^2 r^2} - {(d-1)(d-2) \over 2\kappa^2}
H^2 - \rho \\
\label{R9c}
0&=&{1 \over \kappa^2}\left\{-{k \over r^4} 
 + {(d-2)H_{,t} \over r^2} \right\}
 - (\rho + p) \ .
\eea

When $d=4$, Eqs.(\ref{R8b}) and (\ref{R9b}) have 
the following forms:
\bea
\label{R10}
0&=&-{3k \over 2\kappa^2 r^2} - {3 \over \kappa^2}
H^2 - \rho \nn
&& + (3a + b + c)\left( 6H_{,t}^2 -12 H H_{,tt} 
- 36 H^2 H_{,t} + 6 H^4 \right. \nn
&& \left. + {6kH^2 \over r^2} - {3k^2 \over 2r^4}\right) \ ,\\
\label{R11}
0&=&{1 \over \kappa^2}\left\{{k \over 2r^4} 
 + {2H_{,t} \over r^2} 
 + 3H^2 \right\} - p \nn
&& + (3a + b + c)\left\{ 4 H_{,ttt} + 24 H H_{,tt} + 18 H_{,t}^2 
+ 36 H^2H_{,t} \right. \nn
&& \left. + {k \over r^2}\left( -4 H_{,t} -2 H^2\right)
 + {3k^2 \over 2r^4}\right\}\ .
\eea
Especially if we choose the combination of $R^2$ terms 
in (\ref{vi}) to be proportional to the Gauss-Bonnet 
invariant $G$ or to the square of the Weyl tensor $F$
\bea
\label{R12}
G&=&R^2 -4 R_{\mu\nu}R^{\mu\nu} 
+ R_{\mu\nu\rho\sigma}R^{\mu\nu\rho\sigma}\ ,\nn
F&=&{1 \over 3}R^2 -2 R_{\mu\nu}R^{\mu\nu}
+ R_{\mu\nu\rho\sigma}R^{\mu\nu\rho\sigma}\ ,
\eea
that is, $a=c$, $b=-4a$ or $a={c \over 3}$, $b=-2a$, the 
coefficient $3a + b + c$ vanishes, therefore  $R^2$ 
terms do not give any contribution. As we mentioned first, 
in the Brane New World scenario, $R^2$ terms are produced by 
the trace anomaly, therefore  $R^2$ terms are given by 
a combination of the Gauss-Bonnet invariant and the square 
of the Weyl tensor, therefore, they do not give any contribution. 
The situation is true if we start with the bulk (5d) 
$R^2$-gravity, the holographic trace anomaly is the combination 
of the Gauss-Bonnet invariant and the square of the Weyl 
tensor \cite{anom}. 
Therefore it would be natural that the form  of the FRW 
equation itself does not change even if we start with the bulk 
$R^2$-gravity \cite{SSS}. 

Hence, we reviewed $d$ dimensional FRW dynamics from HD gravity.
The corresponding equations will be used in the next sections.

\section{Boundary terms in bulk $R^2$-gravity}

In section 2, $R^2$-gravity has been considered as an 
effective theory on the $d$ dimensional brane in the 
$d+1$ dimensional bulk space. In this section, we consider 
$d+1$ dimensional $R^2$-gravity  as the gravity in the bulk AdS 
spacetime. Bulk $R^2$-gravity can appear as the next-to-leading 
$1/N$ correction in AdS/CFT correspondence.  

The problem of surface (boundary) counterterms in 
multidimensional HD gravity already has been 
discussed in \cite{NS}. In this section, we  reconsider 
the counterterms corresponding to the Gibbons-Hawking one 
in the Einstein gravity, in $R^2$-gravity and make the arguments 
more rigorous. 
We start from the general $R^2$ part of the total action 
of $d+1$ dimensional $R^2$-gravity:
\be
\label{i}
S_{R^2}=\int d^{d+1} x \sqrt{-\hat G}\left\{a \hat R^2 
+ b \hat R_{\mu\nu} \hat R^{\mu\nu} 
+ c  \hat R_{\mu\nu\xi\sigma} \hat R^{\mu\nu\xi\sigma}\right\}\ .
\ee
By introducing the auxiliary fields $\hat A$, $\hat B_{\mu\nu}$, 
and $\hat C_{\mu\nu\rho\sigma}$, one can rewrite the action  
(\ref{i}) in the following form:
\bea
\label{ii}
\tilde S_{R^2}&=&\int d^{d+1} x \sqrt{-\hat G}
\left\{a \left(2\hat A \hat R
 - \hat A^2\right) + b \left(2\hat B_{\mu\nu} \hat R^{\mu\nu}
 - \hat B_{\mu\nu} \hat B^{\mu\nu}\right) \right. \nn
&& \left. + c \left(2\hat C_{\mu\nu\xi\sigma} 
\hat R^{\mu\nu\xi\sigma} - 
\hat C_{\mu\nu\xi\sigma} \hat C^{\mu\nu\xi\sigma}\right)\right\}\ .
\eea
Using the equation of the motion
\be
\label{iib}
\hat A=\hat R\ ,\quad \hat B_{\mu\nu}=\hat R_{\mu\nu}\ ,\quad 
\hat C_{\mu\nu\rho\sigma}=\hat R_{\mu\nu\rho\sigma}\ , 
\ee
we find the action (\ref{ii}) is equivalent to (\ref{i}). 
Let us impose a Dirichlet type boundary condition, which 
is consistent with (\ref{iib}), 
$\hat A=\left.\hat R\right|_{\rm at\ the\ boundary}$, 
$\hat B_{\mu\nu}=\left.\hat R_{\mu\nu}\right|_{\rm at\ the\ boundary}$, 
and $\hat C_{\mu\nu\rho\sigma}=\left.\hat R_{\mu\nu\rho\sigma}
\right|_{\rm at\ the\ boundary}$ and 
$\delta \hat A = \delta \hat B_{\mu\nu} 
= \delta \hat C_{\mu\nu\rho\sigma}=0$ 
on the boundary. However, the conditions for $\hat B_{\mu\nu}$ and 
$\hat C_{\mu\nu\rho\sigma}$ are, in general, inconsistent. 
For example, even if $\delta B_{\mu\nu}=0$, we have 
$\delta B_{\mu}^{\ \nu}=\delta \hat G^{\nu\rho}B_{\mu\rho}\neq 0$. 
Then one can impose boundary conditions on the scalar 
quantities:
\be
\label{bc1}
\hat A=\hat B_\mu^{\ \mu}=\hat R\ ,
\quad n^\mu n^\nu \hat B_{\mu\nu} 
= n^\mu n^\nu \hat C_{\mu\rho\nu}^{\ \ \ \rho}
= n^\mu n^\nu \hat R_{\mu\nu} \ .
\ee
and 
\be
\label{bc2}
\delta \hat A=\delta\left(\hat B_\mu^{\ \mu}\right)
=\delta\left( n^\mu n^\nu \hat B_{\mu\nu}\right) 
=\delta\left( n^\mu n^\nu \hat C_{\mu\rho\nu}^{\ \ \ \rho}\right)
=0\ .
\ee
Here $n^\mu$ is a unit vector perpendicular to the boundary. 
The equations given by the variations over other components 
could be automatically satisfied by the coordinate or gauge 
choice, as we will see later. 

Using the  conventions of curvatures in 
(\ref{curv}), one can further rewrite the action (\ref{ii}) 
in the following 
form:
\bea
\label{iii}
\tilde S_{R^2}&=&2 \int_{{\rm surface}}d^d x 
\sqrt{-\hat g}\left(
-\hat \Gamma^\lambda_{\mu\rho}n_\nu 
+ \hat \Gamma^\lambda_{\mu\nu}n_\rho
\right) \nn
&& \times \left(a {\delta^\rho}_\lambda \hat G^{\mu\nu}\hat A 
+ b {\delta^\rho}_\lambda \hat B^{\mu\nu} 
+ c {\hat C_\lambda}^{\ \mu\rho\nu}
\right)\nn
&& + \int d^{d+1}\left[\cdots\right]\ .
\eea
Here  
 $\hat g_{mn}$ is the boundary metric induced 
by $\hat G_{\mu\nu}$. 
Now the bulk part of the action denoted by $[\cdots]$ 
does not contain the second order derivative of $\hat G_{\mu\nu}$. 
Then the variational principle becomes well-defined if 
we add the following boundary term to the Einstein action: 
\bea
\label{iv}
\tilde S_{{\rm bndry}}&=&-2 \int_{{\rm surface}}d^d x 
\sqrt{-\hat g}\left(-\hat \Gamma^\lambda_{\mu\nu}n_\nu 
+ \hat\Gamma^\lambda_{\mu\nu}n_\rho
\right) \nn
&&\times \left(a {\delta^\rho}_\lambda g^{\mu\nu}\hat A 
+ b {\delta^\rho}_\lambda \hat B^{\mu\nu} 
+ c {\hat C_\lambda}^{\mu\rho\nu}
\right)\ .
\eea
The action (\ref{iv}) breaks the general covariance. 
We should note, however, that
\be
\label{E4}
\nabla_\mu n_\nu=\partial_\mu n_\nu
 - \hat\Gamma_{\mu\nu}^\lambda n_\lambda \ ,\quad 
\nabla_\mu n^\nu=\partial_\mu n^\nu
 + \hat\Gamma_{\mu\lambda}^\nu n^\lambda \ .
\ee
Then at least for the following metric
\be
\label{E7}
ds^2 = \left(1 + {\cal O}(q^2)\right)dq^2 
+ \hat g_{mn}(q,x^m) dx^m dx^n\ .
\ee
one can write the boundary action (\ref{iv}) as 
\bea
\label{v}
\hat S_{R^2\ {\rm bndry}}&=&\int d^dx \sqrt{-\hat g}\left[
4a\nabla_\mu n^\mu \hat A + 2b\left( n_\mu n_\nu \nabla^\lambda 
n_\lambda + \nabla_\mu n_\nu\right) \hat B^{\mu\nu} \right. \nn
&& \left. + 4cn_\sigma n_\rho \nabla_\mu n_\nu 
\hat C^{\sigma\mu\rho\nu}
\right]\ .
\eea

Choosing $d+1=5$, we start with the 
following bulk action:
\bea
\label{vib}
S&=&\int d^5 x \sqrt{-\hat G}\left\{a \left(2\hat A\hat R - 
\hat A^2\right) + b \left(2\hat B_{\mu\nu} \hat R^{\mu\nu}
 - \hat B_{\mu\nu} \hat B^{\mu\nu}\right) \right. \nn
&& \left. + c \left(2\hat C_{\mu\nu\xi\sigma} 
\hat R^{\mu\nu\xi\sigma}
 - \hat C_{\mu\nu\xi\sigma} \hat C^{\mu\nu\xi\sigma}\right)
+ {1 \over \kappa^2} \hat R - \Lambda \right\}\ .
\eea
We also add the surface terms $S_b^{(1)}$ corresponding to 
Gibbons-Hawking surface term and  (\ref{v}) as well as $S_b^{(2)}$ 
which is the leading 
counterterm corresponding to the vacuum energy on the brane:
\bea
\label{Iiv}
S_b&=&S_b^{(1)} + S_b^{(2)} \nn
S_b^{(1)} &=& \int d^4 x \sqrt{-\hat g}\left[
4 a \nabla_\mu n^\mu \hat A + 2 b\left(n_\mu n_\nu 
\nabla_\sigma n^\sigma
 + \nabla_\mu n_\nu \right) \hat B^{\mu\nu} \right. \nn
&& \left. + 8 c n_\mu n_\nu \nabla_\tau n_\sigma 
\hat C^{\mu\tau\nu\sigma} 
+ {2 \over \kappa^2}\nabla_\mu n^\mu \right] \nn
S_b^{(2)} &=& - \eta\int d^4 x \sqrt{-\hat g} \ .
\eea
Here $\eta$ is a constant, which is determined later. 
We now take the metric of five dimensional space time as follows:
\be
\label{metric1}
ds^2=dz^2 + \e^{2A(z,\sigma)}\sum_{i,j=1}^4
\tilde g_{ij}dx^i dx^j\ ,
\quad \tilde g_{\mu\nu}dx^\mu dx^\nu\equiv l^2\left(d \sigma^2 
+ d\Omega^2_3\right)\ .
\ee
Under the choice of metric in (\ref{metric1}), the curvature 
components are:
\bea
\label{crvtrs}
\hat R_{zi zj}&=&\e^{2A}\left(-A_{,zz} - 
\left(A_{,z}\right)^2\right)\tilde g_{ij} \nn
\hat R_{zA\sigma B}&=&-l^2\e^{2A}A_{,z\sigma}g^s_{AB} \nn
\hat R_{\sigma A\sigma B}&=&\left(-l^2 \e^{2A}A_{,\sigma\sigma} 
-l^2 \e^{4A}\left(A_{,z}\right)^2\right)g^s_{AB} \nn
\hat R_{ABCD}&=&\left(l^2 \e^{2A} -l^2 \e^{2A}
\left(A_{,\sigma}\right)^2
- l^4\e^{4A}\left(A_{,z}\right)^2\right)
\left(g^s_{AC}g^s_{BD} - g^s_{AD}g^s_{BC}\right) \nn
\hat R_{zz}&=&4\left(-A_{,zz} - \left(A_{,z}\right)^2\right) \nn
\hat R_{z\sigma}&=&-3A_{,z\sigma} \nn
\hat R_{\sigma\sigma}&=&l^2\e^{2A}\left(-A_{,zz} 
- 4\left(A_{,z}\right)^2\right) - 3A_{,\sigma\sigma} \nn
\hat R_{AB}&=&\left( l^2\e^{2A}\left(-A_{,zz} 
- 4\left(A_{,z}\right)^2\right) - A_{,\sigma\sigma} 
- 2\left(A_{,\sigma}\right)^2 +2\right) g^s_{AB} \nn
\hat R&=&-8A_{,zz} - 20\left(A_{,z}\right)^2 +l^{-2}\e^{-2A}
\left(-6A_{,\sigma\sigma} - 6\left(A_{,\sigma}\right)^2 +6 
\right) \ .
\eea
Here $\cdot_{,\mu\nu\cdots }\equiv 
{\partial \over \partial x^\mu}{\partial \over \partial x^\nu}
\cdots (\cdot)$. We also write the metric 
of S$_3$ in the following form:
\be
\label{S3m}
d\Omega^2_3=\sum_{A,B=1}^3g^s_{AB}dx^A dx^B\ .
\ee
One gets that $n^\mu$ and the covariant derivative of $n^\mu$ 
are
\be
\label{Dn}
n^\mu=\delta_{\rho}^\mu\ ,\quad 
\nabla_i n^j = \delta_i^j A_{,z}\ (\mbox{others}=0)\ .
\ee
Then the action $S_b$ (\ref{Iiv}) 
looks like:
\bea
\label{SbA}
S_b &=& l^4 \int d^4x \e^{4A}\sqrt{g^s}\left[\left(
16 a \hat A  + 2 b \left(4\hat B_{zz}+\hat B_i^{\ i}\right)  
+ 8c \hat C_{ziz}^{\ \ \ i} \right.\right. \nn
&& \left. \left.+ {8 \over \kappa^2}\right) A_{,z}
+ \eta\right]\ .
\eea
We should note that only the components of $\hat A$, 
$\hat B_{\mu\nu}$, and $\hat C_{\mu\nu\rho\sigma}$ 
in (\ref{bc1}) or (\ref{bc2}) appeared in (\ref{SbA}) since 
$\hat B_{zz} =n^\mu n^\nu \hat B_{\mu\nu}$, 
$\hat B_i^{\ i}=B_\mu^{\ \mu}-n^\mu n^\nu \hat B_{\mu\nu}$, and 
$\hat C_{ziz}^{\ \ \ i}=n^\mu n^\nu 
\hat C_{\mu\rho\nu}^{\ \ \ \rho}$. 
Therefore we need not to consider the variation of these 
auxiliary fields on the boundary. 

 From the variation over $A$, one obtains the following equation 
on the brane, which lies at $z=z_0$:
\bea
\label{brS}
\lefteqn{\left.\delta S\right|_{z=z_0}= 
2V_3l^4 \int d\sigma \sqrt{g^s}\e^{4A} }\nn 
&& \times\left[
\left\{-16a\hat A -2b \left(4 \hat B_{zz} + \hat B_i^{\ i}\right) 
 -8c \hat C_{ziz}^{\ \ \ i}-{8 \over \kappa^2}\right\}
 \delta A_{,z} \right. \nn
&& + \left\{16a\hat A_{,z}
+2b \left(4 \hat B_{zz,z} + \hat B_{i\ ,z}^{\ i}\right) 
+8c \hat C_{ziz\ ,z}^{\ \ \ i} \right. \nn
&& + \left( -16 a\hat A + b \left( 16 \hat B_{zz}
 - 8 \hat B_i^{\ i}\right) \right. \nn
&& \left.\left.\left. + c \left(16 C_{zizi}^{\ \ \ i} 
 - 8 C_{ij}^{\ \ ij}\right)  + {8 \over \kappa^2}\right)A_{,z}
\right\}\delta A\right]\ , \\
\label{brSb}
\lefteqn{\left.\delta S_b\right|_{z=z_0}= 
V_3l^4 \int d\sigma \sqrt{g^s}\e^{4A}\left[\left\{
16 a \hat A  + 2 b \left(4B_{zz} + \hat B_i^{\ i}\right)  
+ 8c \hat C_{ziz}^{\ \ \ i} 
+ {8 \over \kappa^2}\right\} \delta A_{,z} \right.} \nn
&& \left. + 4\left\{ \left( 16 a \hat A  + 2 b 
\left(4B_{zz} + \hat B_i^{\ i} \right) 
+ 8c \hat C_{ziz}^{\ \ \ i} - {8 \over \kappa^2}\right) A_{,z}
+ \eta\right\}\delta A\right]\ .
\eea
Therefore  
\bea
\label{breq}
\lefteqn{\left.\delta\left(S + 2S_b\right)\right|_{z=z_0}} \nn
&=& 2V_3l^4 \int d\sigma \sqrt{g^s}\e^{4A}\left[\left\{
16a\hat A_{,z}
+2b \left(4 \hat B_{zz,z} + \hat B_{i\ ,z}^{\ i}\right) 
+8c \hat C_{ziz\ ,z}^{\ \ \ i} \right.\right. \nn
&& \left.\left. + \left( 48 a\hat A + 48b \hat B_{zz}
 + c \left(48 C_{ziz}^{\ \ \ i} 
 - 8 C_{ij}^{\ \ ij}\right)+{24 \over \kappa^2} 
\right)A_{,z} + 4\eta \right\}\delta A\right]\ .
\eea
The factors $2$ in front of $S_b$ and $V_3$ come from the fact 
that we are considering two bulk spaces, which have one 
common boundary or brane. $V_3$ is the volume of the unit 3 sphere:
\be
\label{V3}
V_3=\int d^3x_A\sqrt{g^s}=2\pi^2\ .
\ee
Since the terms containing $\delta A_{,z}$ do not appear, 
the variational principle is well-defined. 
Then one obtains the following equation on the boundary or 
brane 
\bea
\label{viii}
0&=&16a\hat A_{,z}
+2b \left(4 \hat B_{zz,z} + \hat B_{i\ ,z}^{\ i}\right) 
+8c \hat C_{ziz\ ,z}^{\ \ \ i}  \nn
&& + \left( 48 a\hat A + 48b \hat B_{zz}
 + c \left(48 C_{ziz}^{\ \ \ i} 
 - 8 C_{ij}^{\ \ ij}\right)+{24 \over \kappa^2} 
\right)A_{,z} + 4\eta \nn
&=&16aR_{,z} +2b \left(4R_{zz,z} + R_{i\ ,z}^{\ i}\right) 
+8c \hat C_{ziz\ ,z}^{\ \ \ i} \nn
&& + \left( 48 a\hat A + 48b \hat B_{zz}
 + c \left(56 R_{zz} - 8 R_{i}^{\ i}\right)+{24 \over \kappa^2} 
\right)A_{,z} + 4\eta \ .
\eea
In the second line, (\ref{iib}) is used. 
Especially taking the background as AdS-Schwarzschild 
black hole, one gets 
\be
\label{cv}
\hat R=-{20 \over l^2}\ ,\quad 
\hat R_{\mu\nu}= - {4 \over l^2}g_{\mu\nu}\ .
\ee
Here $l$ is the radius of the asymptotic AdS space, given by 
solving the equation
\be
\label{ll}
0={80 a \over l^4} + {16 b \over l^4} + {8c \over l^4}
- {12 \over \kappa^2 l^2}-\Lambda\ .
\ee
Then from (\ref{viii}), we have
\be
\label{ix}
0=-3\left({320 a \over l^2} + {64b \over l^2} 
+ {32c \over l^2} -{8 \over \kappa^2} \right) 
A_{,z} + 4\eta \ .
\ee
The parameter $\eta$ is not the free parameter. It can be 
determined from the condition that the leading order 
divergence of the action, which appears when the brane goes 
to infinity in the asymptotic AdS space, is cancelled. 
If we consider the asymptotic anti de Sitter space:
\bea
\label{SAdS0}
ds^2&\sim&-\e^{2\rho_0}dt^2 + \e^{-2\rho_0}dr^2 
+ r^2\sum_{i,j}^3 g_{ij}dx^i dx^j\ ,\nn
\e^{2\rho_0}&=&{1 \over r^2}\left({kr^2 \over 2} 
+ {r^4 \over l^2}\right)\ ,
\eea
one finds 
\bea
\label{clactions}
S&=&\int d^4x r_0^4 {1 \over 4} \left\{{400 a \over l^4}
+ {80b \over l^4} + {40c \over l^4} - {20 \over \kappa^2 l^2}
 -\Lambda \right\} + {\cal O}\left(r_0^3\right) \nn
&=&\int d^4x r_0^4 \left\{{80 a \over l^4}
+ {16b \over l^4} + {8c \over l^4} - {2 \over \kappa^2 l^2}
\right\} + {\cal O}\left(r_0^3\right) \nn
S_b&=&-\int d^4x r_0^d \left\{ {320 a \over l^2} 
+ {64b \over l^2} + {32c \over l^2} - {8 \over l^2 \kappa^2} 
 - {\eta \over l}\right\} + {\cal O}\left(r_0^3\right) \ .
\eea
Here we assume there is a brane at $r=r_0$. 
In the second line, we have deleted $\Lambda$ by 
using (\ref{ll}). 
Then one gets
\be
\label{eta1}
{\eta \over l}={240 a \over l^4} + {48b \over l^4}
+ {24c \over l^4} - {6 \over \kappa^2 l^2}\ .
\ee
Then combining Eq.(\ref{ix}) and (\ref{eta1}), one obtains a 
simple equation  in the form which appeared in the previous 
papers\cite{NOZ,SSS} 
\be
\label{x}
0=A_{,z} - {1 \over l}\ .
\ee
As a special case, we can consider the case that the $R^2$ terms 
are given by the combination of the Gauss-Bonnet term (\ref{R12}) 
in four dimensions:
\be
\label{GBcmb}
b=-4a\ ,\quad c=a\ .
\ee
Then Eqs.(\ref{ll}), (\ref{ix}) and (\ref{eta1}) have the 
following forms:
\bea
\label{llb}
0&=&{24 a \over l^4} - {12 \over \kappa^2 l^2}-\Lambda\ ,\\
\label{ixb}
0&=&-3\left({96 a \over l^2} -{8 \over \kappa^2} \right) 
A_{,z} + 4\eta \ ,\\
\label{eta1b}
{\eta \over l}&=&{72 a \over l^4} - {6 \over \kappa^2 l^2}\ .
\eea
Hence, the surface counterterms (which are useful for regularization of 
thermodynamical quantities) in bulk HD gravity 
around asymptotically AdS space are derived. These counterterms will 
be used later on when the discussion of
 FRW-dynamics induced by brane will be presented.

\section{Thermodynamics and entropy of bulk AdS black hole}

In this section we will be interested in thermodynamics of AdS BH 
with non-trivial boundary (brane) in bulk $R^2$-gravity.
The case is considered when HD terms contain the Riemann 
tensor square term, i.e. 
$R_{\mu\nu\xi\sigma}R^{\mu\nu\xi\sigma}$. 
The calculation of thermodynamical quantities like mass and entropy 
will be necessary in order to relate them with the corresponding 
ones in brane FRW Universe. In section 5, by using the brane 
equation (\ref{x}) obtained in section 3, we find that the 
dynamics of the brane is given by the FRW-like equation. From the 
FRW-like equation, we can find the thermodynamical quantities and 
we will compare the quantities with those obtained in this section. 

The general action of $d+1$ dimensional $R^2$-gravity is given 
by (\ref{vi}).
When $c=0$, Schwarzschild-anti de Sitter space is an exact solution:
\bea
\label{SAdS}
ds^2&=&\hat G_{\mu\nu}dx^\mu dx^\nu \nn
&=&-\e^{2\rho_0}dt^2 + \e^{-2\rho_0}dr^2 
+ r^2\sum_{i,j}^{d-1} g_{ij}dx^i dx^j\ ,\nn
\e^{2\rho_0}&=&{1 \over r^{d-2}}\left(-\mu + {kr^{d-2} \over d-2} 
+ {r^d \over l^2}\right)\ .
\eea
For non-vanishing $c$, such an S-AdS BH solution may be
constructed perturbatively\cite{SNO}. In this case, it is useful to 
establish the higher-derivative AdS/CFT correspondence\cite{NS1} 
and find the strong coupling limit of super Yang-Mills theory 
with two supersymmetries in next-to-leading order.
In this section, we only consider the case $a=b=0$ 
(see also Appendix) for simplicity: 
\be
\label{rie}
S=\int d^{d+1} x \sqrt{-\hat G}\left\{
c \hat R_{\mu\nu\xi\sigma}\hat R^{\mu\nu\xi\sigma}
+ {1 \over \kappa^2} \hat R - \Lambda \right\}\ .
\ee
When we assume the metric (\ref{SAdS}) with $\mu=0$, the scalar, 
Ricci and Riemann curvatures are given by
\bea
\label{rr}
\hat{R}=- {d(d+1)\over l^2},\; \hat{R}_{\mu\nu}=
-{d\over l^{2}}G_{\mu\nu},\; \hat{R}_{\mu\nu\xi\sigma} 
= -{1\over l^2}\left( \hat{G}_{\mu\xi}\hat{G}
_{\nu\sigma} -\hat{G}_{\mu\sigma}\hat{G}_{\nu\xi} \right), 
\eea
which tell that the curvatures are covariantly constant. The 
equation of the motion derived from the action (\ref{rie})
is:
\bea
\label{cl}
0 &=& -{1\over 2}\hat{G}_{\zeta\xi}\left\{ c \hat{R}_{\mu\nu\rho\sigma}
\hat{R}^{\mu\nu\rho\sigma} +{1\over \kappa^{2}}\hat{R}-\Lambda \right\}  \nn
&& + 2c \hat{R}_{\zeta\mu\nu\rho}\hat{R}_{\xi}^{\mu\nu\rho}
+{1\over \kappa^{2}}\hat{R}_{\zeta \xi}+4c D_{\rho}D_{\kappa}\hat{R}_{\zeta\; \xi}^{\; \rho \; \kappa}.
\eea
Then substituting Eqs.(\ref{rr}) into (\ref{cl}), one finds the
relation between $c$, $\Lambda$ and $l$
\bea
\label{clc}
0={2c \over l^4}d(d-3)-{d(d-1) \over \kappa^{2} l^{2}}-\Lambda\ ,
\eea 
which defines the radius $l$ of the asymptotic AdS space 
even if $\mu\neq 0$. 
For $d+1=5$ with $\mu\neq 0$,  using Eq.(\ref{cl}),  
we get the perturbative solution from (\ref{SAdS}), 
which looks like \cite{NS1}:
\bea
\label{dAdS}
\e^{2\rho}={1\over r^2}\left\{ -\mu +{k\over 2}r^2 + {r^{4}\over l^2}
+{2\mu^2 \epsilon \over r^{4} }\right\}, \quad 
\epsilon = c\kappa^{2}\ .
\eea
Suppose that $g_{ij}$ (\ref{SAdS}) corresponds to the Einstein 
manifold, 
defined by $r_{ij}=kg_{ij}$, where $r_{ij}$ is Ricci tensor 
defined by $g_{ij}$ and $k$ is the constant. 
For example, if $k>0$ the boundary can be three dimensional 
sphere, if $k<0$, hyperboloid, or if $k=0$, flat space. 
Properly normalizing the coordinates, one can choose $k=2$, $0$, 
or $-2$. Note that as it will be shown in next section dual QFT 
corresponding to above background is not CFT. Probably, the fact 
that there is no exact S-AdS solution of non-zero c HD gravity 
is manifestation of this property. 

The calculation of thermodynamical quantites like free 
energy $F$, the entropy ${\cal S}$ and the energy $E$ may be done 
following \cite{NS1}. After Wick-rotating the time variable by 
$t \to i\tau$, the
free energy $F$ can be obtained from the action $S$ (\ref{rie}), 
where the classical solution is substituted:
\bea
F=-TS\ .
\eea
Multiplying $\hat{G}^{\zeta\xi}$ to (\ref{cl}) in case 
that $D_{\rho}D_{\kappa}
\hat{R}_{\zeta\; \xi}^{\; \rho \; \kappa}={\cal O}(\epsilon)$ 
as in the solution (\ref{dAdS}), we find for $d=4$
\be
\label{cl2}
{1\over \kappa^{2}}\hat{R}=- {c \over 3} \hat{R}_{\mu\nu\rho\sigma}
\hat{R}^{\mu\nu\rho\sigma} + {5 \over 3}\Lambda + 
{\cal O}\left(\epsilon^2\right)\ .
\ee
Substituting (\ref{cl2}) into the action (\ref{rie}), 
one arrives to the following expression:
\be
\label{ri2}
S=\int d^5 x \sqrt{-\hat G}\left\{
{2 \over 3}c \hat R_{\mu\nu\xi\sigma}\hat R^{\mu\nu\xi\sigma}
+ {2 \over 3} \Lambda \right\}\ .
\ee
Since 
\be
\label{riesqr}
\hat R_{\mu\nu\xi\sigma}\hat R^{\mu\nu\xi\sigma}
={40 \over l^2}+{72 \mu^2 \over r^8} + {\cal O}(\epsilon)\ ,
\ee
and using (\ref{clc}) with $d=4$,  we obtain
\bea
\label{sr}
S &=& - \int d^{5}x \sqrt{-\hat G} \left( {8 \over \kappa^2 l^{2}}
 - {32 c \over l^4} + {48c \mu^2 \over r^8}\right) \nn
&=& -{V_{3} \over T}\int ^{\infty}_{r_{H}} dr r^{3}
\left( {8 \over \kappa^2 l^{2} }- {32 c \over l^4} 
 + {48c \mu^2 \over r^8}\right)\ .
\eea
Here $V_{3}$ is the volume of 3d sphere and we assume $\tau$ 
has a period  ${1\over T}$.  The expression for $S$ contains 
the divergence coming from large $r$. In order to subtract the 
divergence, we regularize $S$ (\ref{sr}) by cutting off the 
integral at a large radius $r_{\rm max}$ and subtracting the 
solution with $\mu =0$ in a same way as in \cite{NS}: 
\bea
S_{\rm reg}&=& 
-{V_{3} \over T}\left\{ \int ^{r_{\rm max}}_{r_{H}} dr r^{3}
\left( {8 \over \kappa^2 l^{2} }- {32 c \over l^4} 
+ {48c \mu^2 \over r^8}\right) \right. \nn
&& \left. -\e^{\rho(r=r_{\rm max})-\rho(r=r_{\rm max};\mu =0) }
\int ^{r_{\rm max}}_{0} dr r^{3} 
\left( {8 \over \kappa^2 l^{2} }- {32 c \over l^4} \right)
\right\}\ .
\eea
The factor $\e^{\rho(r=r_{\rm max})-\rho(r=r_{\rm max};\mu =0)}$ 
is chosen so that the proper length of the circle which 
corresponds to the period ${1 \over T}$ in the Euclidean 
time at $r=r_{\rm max}$ coincides with each other in the 
two solutions.  Taking $r_{\rm max} \to \infty$, one finds 
\bea
F= V_{3}\left\{\left( {l^2 \mu \over 8} 
 - {r_H^4  \over 4}\right)
\left( {8 \over \kappa^2 l^{2} }- {32 c \over l^4} \right)
+ {12 c\mu^2 \over r_H^4}\right\}\ .
\eea
The horizon radius $r_{H}$ is given by solving 
the equation $\e^{2\rho_0(r_H)}=0$ in (\ref{dAdS}). 
We can solve $r_{H}$ perturbatively up to first order 
on $c$ by putting $r_{H}=r_{0}+ c\delta r$, 
where $r_{0}$ is the horizon radius when $c=0$. 
As in \cite{SSS}, $r_{0}$ is given by 
\bea
\label{rh1}
r_{0}^{2}=-{k l^{2} \over 4} + {1\over 2}
\sqrt{{k^2 \over 4}l^{4}+ 4\mu l^{2} } \; .
\eea
Then the horizon radius $r_{H}$ is obtained as follows;
\bea
\label{hrznrds}
r_{H}=r_{0}-{c\mu^{2}\kappa^{2} \over r_{0}^{3}
\left( 2\mu -{k \over 2}r_{0}^{2} \right) }\ .
\eea
We can also rewrite the black hole mass $\mu$ (using $r_{H}$)
up to first order on $\epsilon$, $(\epsilon = c\kappa^{2})$.
\bea
\label{mass}
\mu &=& {k\over 2}r_{H}^{2}+{r_{H}^{4} \over l^{2}} 
+{2 \epsilon \over r_{H}^{4}}\left( {k \over 2}r_{H}^{2}
+{r_{H}^{4} \over l^{2} } \right)^{2}\ .
\eea 
Then  $F$ looks like
\bea
F= {V_{3} \over \kappa^{2} l^{2}} \left[ {l^{2} k \over 2}
r_{H}^{2} -r_{H}^{4}+\epsilon \left\{ {l^{2}k^{2} \over 2}
+{6r_{H}^{4}\over l^{2} } + {12 l^2 \over r_H^4}
\left( {k \over 2}r_{H}^{2}
+{r_{H}^{4} \over l^{2} } \right)^{2} \right\}\right]\ .
\eea
The Hawking temperature $T_H$ is given by 
\bea
\label{ht1}
T_H &=& {(\e^{2\rho})'|_{r=r_{H}} \over 4\pi} \\
&=& {1 \over 4\pi}\left\{ {4r_{H} \over l^2 }+ {k\over r_{H}}
-{8\epsilon \over r_{H}^{7} } \left({k\over 2}r_{H}^{2}
+ {r_{H}^{4} \over l^2} 
\right)^{2} \right\} \; , \nonumber
\eea
where $'$ denotes the derivative with respect to $r$.
Then the entropy ${\cal S}$ and energy $E$ have the following form:  
\bea
\label{entropy}
{\cal S }&=& -{dF \over dT_H}
=-{dF \over dr_{H}}{dr_{H} \over dT_H} \nn
&=& {4\pi V_{3} r_{H}^{3} \over \kappa^{2} }
\left\{ 1- {16 \epsilon \over l^{2}} 
\left( 1+{k l^{2} \over 2 r_{H}^{2} } 
+ {3 k^{2} l^{4} \over 32 r_{H}^{4} } \right) 
\left( 1-{k l^{2} \over 4 r_{H}^{2} } \right)^{-1} 
\right\} \ ,\\
\label{energy}
E &=& F+T{\cal S} \nn
&=& {3 V_{3} \over \kappa^{2} }
\left\{ {1\over 2}kr_{H}^{2}+{r_{H}^{4} \over l^{2}} \right. 
\left. -\epsilon \left( {18 r_{H}^{4} \over l^{4} }
+{31 k r_{H}^{2} \over 2 l^{2}}
+{9 \over 2}k^{2}+{5 k^{3} l^{2} \over 8 r_{H}^{2}} \right)
\left( 1-{k l^{2} \over 4 r_{H}^{2} } \right)^{-1}
\right\} \nn
&=& {3 V_{3} \over \kappa^{2} }\left[ \mu 
- {\epsilon \over l^2}
\left\{k^3 l^4 + 8 k l^2 \mu 
+ \left(k^2 l^2 + 20 \mu \right)\sqrt{k^2 l^4 + 16 l^2\mu}
\right\} \right.\nn
&& \left.\times\left(-2 k l^2 + \sqrt{k^2 l^4
 + 16 l^2\mu}\right)^{-1}\right]\ .
\eea
Hence, we described the thermodynamics of AdS BH where observable 
Universe may appear as corresponding brane. Some remarks 
are in order. It is remarkable that the entropy ${\cal S}$ 
is not proportional to the area of the horizon when $k\neq 0$ 
and the energy $E$ is not to $\mu$, either. We should note that 
the entropy ${\cal S}$ was 
proportional to the area and the energy $E$ to $\mu$ even in 
$R^2$-gravity if there is no the squared Riemann tensor 
term ($c=0$ in (\ref{vib})) \cite{SSS}, where we have 
the following expressions:
\bea
F&=& -{V_{3} \over 8}r_{H}^{2} \left( {r_{H}^{2} \over l^{2}}
 - {k \over 2} \right)\left( {8 \over \kappa^2} 
 - {320 a \over l^2} -{64 b \over l^2} \right) \; , \\
\label{ent}
{\cal S }&=&{V_{3}\pi r_H^3 \over 2}
\left( {8 \over \kappa^2}- {320 a \over l^2}
 -{64 b \over l^2} \right)\ ,\\
\label{ener}
E&=& {3V_{3}\mu \over 8}
\left( {8 \over \kappa^2}- {320 a \over l^2}
 -{64 b \over l^2} \right)\ .
\eea
Here $a$ and $b$ are given in (\ref{vib}). 

Several authors have investigated the entropy of higher derivative 
gravity using the first law of thermodynamics and 
the Noether current \cite{sR}. 
With the help of formula from \cite{sR}, one gets the following expression 
of the entropy ${\cal S}$ corresponding to the action in (\ref{vib}): 
\bea
\label{JKM}
{\cal S}&=&{4\pi \over \kappa^2}\int_{\rm horizon} d^3x 
\sqrt{g_H}
\left\{1 + 2a\kappa^2 \hat R + b\kappa^2 \left(\hat R - g_H^{ij}
\hat R_{ij}\right) \right.\nn
&& \left.+ 2c \left( \hat R  - 2g_H^{ij}\hat R_{ij} 
+ g_H^{ij} g_H^{kl}\hat R_{ikjl}\right)
\right\}\ .
\eea
Here $g_H$ is the metric induced on the horizon and we follow 
the notations in this paper. When $c=0$ and the curvatures 
are given by (\ref{cv}), we have $\hat R =-{20 \over l^2}$ and 
$g_H^{ij}\hat R_{ij}=-{12 \over l^2}$ for $d=4$. Then 
Eq.(\ref{JKM}) has the following form:
\be
\label{ent2}
{\cal S }={V_{3}\pi r_H^3 \over 2}
\left( {8 \over \kappa^2}- {320 a \over l^2} 
  -{64 b \over l^2} \right)\ .
\ee
Therefore the result (\ref{ent}) is exactly reproduced. 
Eq.(\ref{ent}) tells that the entropy is proportional to 
the area of the horizon if $c=0$ but Eq.(\ref{entropy}) seems 
to tell that the entropy is not when $c\neq 0$. 
When $a=b=0$ and $c\neq 0$ one gets
\bea
\label{hrznR}
g_H^{ij} g_H^{kl}\hat R_{ikjl}&=&
 -{6 \over l^2} + {6\mu \over r_H^4} + {\cal O}(\epsilon) \nn
&=& {3k \over r_H^2} + {\cal O}(\epsilon) \ .
\eea
Then 
\bea
\label{JKMentropy}
{\cal S }&=& {4\pi V_{3} r_{H}^{3} \over \kappa^{2} }
\left\{ 1 + {\epsilon \over l^{2}} 
\left( 8 + {6k l^{2} \over r_{H}^{2} }\right) \right\} \ ,
\eea
which does not agree with (\ref{entropy}). 
We should note, however, that, since we define ${\cal S}$ in 
(\ref{entropy}) and $E$ in (\ref{energy}) based on thermodynamics, 
it is clear that these quantities satisfy the first law:
\be
\label{1thr}
T\delta{\cal S}=\delta E\ ,
\ee
for static (not rotating) case even if $c\neq 0$. The difference 
between the entropies might express the regularization 
(parametrization) dependence of the entropy. 
Indeed in our calculation, the traditional regularization method is applied.
In the calculation of refs.\cite{sR} thermodynamics is defined on horizon,
so in their case the regularization is implicit. Moreover, they mainly 
considered asymptotically flat backgrounds. Note that the entropy 
could be changed in general by the ambiguity in the choice of 
boundary terms (see section 3) which are necessary to 
make the variational 
principle well-defined and (or) to make the action finite.
However the question remains why there was coincidence of two entropies 
when $c=0$ in HD gravity? The answer is probably given 
in next section where we show that non-zero $c$ HD gravity on 
such S-AdS BH has non-CFT field theory as dual one. 
The breaking of conformal invariance of dual QFT may cause 
the physical reason for such disagreement between entropies.
Note in this connection that presumbly logarithmic CFT (for recent 
discussion, see \cite{IKN})
 should be used to describe dual QFT near $c=0$ barrier.

In order to clarify the situation let us consider the 
formula (\ref{JKM}) for the entropy 
as well as (\ref{entropy}) and (\ref{ent}) when 
$b\neq 0$ and/or $c\neq 0$ but $a\kappa^2$, 
$b\kappa^2$ and $c\kappa^2$ are small. 
Combining (\ref{entropy}) and (\ref{ent}), we have
\bea
\label{entropy2}
{\cal S }&=& {4\pi V_{3} r_{H}^{3} \over \kappa^{2} }
\left\{ 1 - {40 a\kappa^2 \over l^2} -{8 b\kappa^2 \over l^2} 
 - {4c\kappa^2 \over l^2} \right.\nn
&& \left. -{3 c\kappa^2 \over 2l^{2}} 
\left( 4 + {k l^2 \over r_{H}^{2} }\right) 
\left(2 + { k l^2 \over r_{H}^2 } \right) 
\left( 1-{k l^{2} \over 4 r_{H}^{2} } \right)^{-1} 
\right\} \ .
\eea
Note that the combination of the first 4 terms in the first line 
often appeared, for example, in (\ref{ix}) and (\ref{eta1}), it 
might have a universal meaning. Eq.(\ref{entropy2}) can be 
further rewritten in the following form: 
\bea
\label{entropy2b}
{\cal S }&=& {4\pi V_{3} r_{H}^{3} \over \kappa^{2} }
\left\{ 1 - {40 a\kappa^2 \over l^2} -{8 b\kappa^2 \over l^2} 
+ {c\kappa^2 \over l^2}\left(8 + {6kl^2 \over r_H^2}\right) 
\right.\nn
&& \left. - {12c\kappa^2 \over l^2} \left(2 + { k l^2 \over r_{H}^2 } \right) 
\left( 1-{k l^{2} \over 4 r_{H}^{2} } \right)^{-1} 
\right\} \ .
\eea
The last term depends on the curvature of the horizon $k$. 
Since the scalar curvature $R_H$ of the horizon is given by
\be
\label{RH}
R_H={3k \over r_H^2} \ ,
\ee
by comparing (\ref{entropy2b}) with (\ref{JKM}), 
(\ref{ent2}) and (\ref{JKMentropy}), 
we propose the following formula for the entropy:
\bea
\label{gnrlzdJKM}
{\cal S}&=&{4\pi \over \kappa^2}\int_{\rm horizon} d^3x 
\sqrt{g_H}
\left\{1 + 2a\kappa^2 \hat R + b\kappa^2 \left(\hat R - g_H^{ij}
\hat R_{ij}\right) \right. \nn
&& + 2c \left( \hat R  - 2g_H^{ij}\hat R_{ij} 
+ g_H^{ij} g_H^{kl}\hat R_{ikjl}\right) \nn
&& \left. - {12c\kappa^2 \over l^{2}} \left(2 
+ {R_H l^2 \over 3} \right) 
\left( 1-{ R_H l^{2} \over 12 } \right)^{-1} \right\}\ .
\eea
The last term diverges when $R_H={12 \over l^2}$. Since $l$ is 
the length parameter of the 5d AdS space, this occurs when the 
size of the black hole is very large (of cosmological size). We 
do not expect that the perturbation with respect to $c\kappa^2$ 
is valid in this case. Therefore the singularity might not be real 
but apparent one. This consideration shows that small parameter 
expansion as done in this section has only limited region of validity.

Hence we calculated thermodynamical quantities (entropy, free energy, 
Hawking temperature) of S-AdS BH in five dimensional HD gravity. 
BH under consideration contains the brane (four dimensional universe).
Thermodynamical description of this section will be used in 
next section for comparison with FRW-description of brane 
(and dual QFT).

\section{Brane equation as FRW equation}

We now consider the brane equation (\ref{x}) in 
five dimensional bulk space of black hole type. 
Such a brane equation can be regarded as FRW 
equation of the brane world. From FRW equation, we can find 
the energy and the entropy of the matter in the brane universe. 
One can compare these thermodynamical quantities with 
those in section 4, obtained from the bulk black hole. 
Since  $c=0$ case has been investigated in \cite{SSS}, we now 
concentrate on the situation when $a=b=0$ and $c\neq 0$. The 
generalization to the case when all parameters $a,b,c$ are non-zero 
is straitforward and does not change the qualitative conclusions. 

Let us rewrite the metric (\ref{dAdS}) of Schwarzschild-anti 
de Sitter space with correction in a form of (\ref{metric1}). 
If one chooses coordinates $(q,\tau)$ as
\bea
\label{cc1}
&& l^2\e^{2A-2\rho}A_{,q}^2 - \e^{2\rho} t_{,q}^2 = 1 \ ,
\quad l^2\e^{2A-2\rho}A_{,q}A_{,\tau}
 - \e^{2\rho}t_{,q} t_{,\tau}= 0 \ ,\nn
&& l^2\e^{2A-2\rho} A_{,\tau}^2 - \e^{2\rho} t_{,\tau}^2 
= -l^2\e^{2A}\ . 
\eea
the metric takes the form (\ref{metric1}). Here $r=l\e^A$.
Furthermore choosing a coordinate $\tilde t$ by 
$d\tilde t = l\e^A d\tau$, 
the metric on the brane takes FRW form: 
\be
\label{e3}
ds_{\rm brane}^2= -d \tilde t^2  + l^2\e^{2A}
\sum_{i,j=1}^3g_{ij}dx^i dx^j\ .
\ee
Solving Eqs.(\ref{cc1}), we have
\be
\label{e4}
H^2 = A_{,q}^2 - {\e^{2\rho}\e^{-2A} \over l^2}\ .
\ee
Here the Hubble constant $H$ is defined by $H={dA \over d\tilde t}$. 
Then using (\ref{dAdS}),  one obtains the following equation:
\be
\label{e5}
H^2={\mu \over r^4} - {k\over 2r^2} 
- {2\mu^2 \epsilon \over r^8 }\ .
\ee
Especially, when $k=2>0$, the spacial part of the brane has the 
shape of the three dimensional sphere and $r$ can be regarded 
as the radius of the spacial part of the brane universe. 
The last term in (\ref{e5}) is  unusual, it did not  
appear even in the $R^2$-gravity with $c=0$. 

Eq.(\ref{e5}) can be rewritten in the form of the 
FRW equation (compare with \cite{SV}):
\bea
\label{F1}
H^2 &=& - {k \over 2r^2} + {\kappa_4^2 \over 6}
{\tilde E \over V}\ ,\nn
{\tilde E}&=&{6 V_3 \over \kappa_4^2 r}
\left(\mu - {2\epsilon\mu^2 \over r^4}\right)
\ ,\nn
V&=&r^3V_3\ .
\eea
Here $V_3$ is the volume of the three dimensional sphere 
with a unit radius  and $\kappa_4$ is the four dimensional 
gravitational coupling, which is given by
\be
\label{F2}
\kappa_4^2={2 \tilde\kappa^2 \over l}\ ,\quad
{1 \over \tilde \kappa^2}\equiv {1 \over \kappa^2}
 - {4c \over l^2}\ .
\ee
Differentiating Eq.(\ref{F1}) with respect to $\tilde t$, 
since $H={1 \over r}{dr \over d\tilde t}$, we obtain the 
second FRW equation
\bea
\label{2FR1}
\dot H &=& - {\kappa_4^2 \over 4} \left({\tilde E \over V} 
+ p\right) + {k \over 2r^2}\ ,\nn
p&=&{2 \over r^4 \kappa_4^2}\left(\mu
 - {10\epsilon\mu^2 \over r^4}\right)
\ ,
\eea
if $H\neq 0$. We should note that Eq.(\ref{2FR1}) needs not 
to be satisfied for the static solution where $H=0$. 
Here $p$ can be regarded as the pressure of the matter on 
the brane.

Note that when $r$ is large, the metric (\ref{dAdS}) has 
the following form:
\be
\label{eq13} 
ds_{\rm AdS-S}^2 \rightarrow {r^2 \over l^2}\left(-dt^2 
+ l^2 \sum_{i,j}^3 
g_{ij}dx^i dx^j\right)\ ,
\ee
which tells that  CFT time $\tilde t$ is equal 
to  AdS time $t$ times the factor ${r \over l}$:
\be
\label{eq14}
t_{\rm CFT}={r \over l}t\ .
\ee
Therefore the energy $\tilde E$ (\ref{F1}) in CFT should be related 
with the energy $E$ (\ref{energy}) in AdS BH by a 
factor ${l \over r}$ \cite{SV}: 
\be
\label{F3}
\tilde E={l \over r}E\ .
\ee
However, Eq.(\ref{F3}) is not satisfied in general. 
Using (\ref{F2}), the energy in (\ref{energy}) can be rewritten as: 
\bea
\label{energyAN}
E&=&{6 V_{3} \over \kappa_4^{2} l}\left\{ \mu 
 -  {\epsilon \over l^{2}} \left(k^2 l^2 + 16 \mu\right)\left(k l^2 
+ \sqrt{k^2 l^4 + 16 l^2\mu}\right) \right. \nn
&& \left. \times\left(-2 k l^2 + \sqrt{k^2 l^4 
+ 16 l^2\mu}\right)^{-1}\right\}\ ,
\eea
Then comparing (\ref{energyAN}) with (\ref{F1}), we find that 
Eq.(\ref{F3}) is satisfied when the brane exists at $r=r_{\rm br}$:
\bea
\label{bbrs1}
{1 \over r_{\rm br}^4}&=&{1 \over 2l^2\mu^2 }{
\left(k^2 l^2 + 16 \mu\right)\left(k l^2 
+ \sqrt{k^2 l^4 + 16 l^2\mu}\right) 
\over \left(-2 k l^2 + \sqrt{k^2 l^4 
+ 16 l^2\mu}\right)} \nn
&=&{4\left( 4 r_{H}^2 + kl^2\right)^2 \over r_H^4 
\left(4r_H^2 - kl^2\right)\left(2r_H^2 -kl^2\right)}\ .
\eea
 From Eqs.(\ref{F1}) and (\ref{2FR1}), one gets 
\be
\label{trace1}
-{\tilde E \over V} + 3p= - {48\epsilon \mu^2 \over 
\kappa_4^2 r^8}\ ,
\ee
which tells that the trace of the energy-stress tensor coming 
from the matter on the brane does not vanish:
\be
\label{trace2}
{T^{{\rm matter}\ \mu}}_\mu\neq 0\ .
\ee
Therefore the conformal symmetry of the matter on the brane 
should be broken by the correction coming from the squared 
Riemann tensor term. The conformal symmetry could be restored 
when $r\rightarrow \infty$. 

In \cite{EV}, it was shown that FRW equation in 
$d$ dimensions can be regarded as a $d$ dimensional analogue of 
the Cardy formula of 2d conformal field theory (CFT) \cite{Cardy}:
\be
\label{CV1}
\tilde {\cal S}=2\pi \sqrt{
{c \over 6}\left(L_0 - {k \over d-2}{c \over 24}\right)}\ .
\ee
In the present case, identifying 
\bea
\label{CV2}
{2\pi \tilde E r \over d-1} &\Rightarrow& 2\pi L_0 \ ,\nn
{(d-2)V \over \kappa_d^2 r} &\Rightarrow& {c \over 24} \ ,\nn
{4\pi (d-2)HV \over \kappa_d^2} &\Rightarrow& \tilde {\cal S}\ ,
\eea
the FRW-like equation (\ref{F1}) has the same form as (\ref{CV1}) 
(for related discussion of this formula see papers \cite{others}). 

The total entropy of the universe could be conserved during the 
expansion. Then one can evaluate holographic (Hubble) entropy 
$\tilde{\cal S}$ in (\ref{CV2}) when the brane crosses the 
horizon $r=r_H$. When $r=r_H$, Eq.(\ref{F1}) tells that 
\be
\label{CV3}
H=\pm {1 \over l}\ .
\ee
Here the plus sign corresponds to the expanding brane universe 
and the minus one to the contracting universe. Taking 
the expanding case and using (\ref{CV2}), we find
\be
\label{CV4}
\tilde{\cal S}={4\pi (d-2) V \over l\kappa_d^2}
={2\pi (d-2) r_H^{d-1} V_{d-1} \over \tilde\kappa^2}\ .
\ee
When $c=0$ but $a$, $b\neq 0$ in general, the entropy 
$\tilde{\cal S}$ is identical with the black hole 
entropy ${\cal S}$. When $c\neq 0$, however, there is 
a difference between the two kinds of entropies. 
For $d=4$,  using (\ref{entropy}) and (\ref{CV4}), 
we find 
\be
\label{CV5}
\tilde{\cal S}-{\cal S}
={6\pi \epsilon V_{3} r_{H}^{3} \over l^2\kappa^{2} }
\left( 4 +{k l^{2} \over r_{H}^{2} } \right)
\left(2 + {k l^2 \over r_{H}^2 } \right) 
\left( 1-{k l^{2} \over 4 r_{H}^{2} } \right)^{-1} \ .
\ee
Formula (\ref{CV1}) has been derived for the conformal field 
theory. Eq.(\ref{trace2}), however, tells that the conformal 
invariance of dual theory is broken. The difference 
between $\tilde{\cal S}$ and ${\cal S}$ in (\ref{CV5}) might 
express the fact that proposed dual QFT is not CFT.
Then, the difference of entropies may serve as some measure 
for deviation from AdS/CFT correspondence. It would be 
interesting to analyze the physics behind this property. 
One possible speculation is that it could help
in formulating of some sort of AdS/non-CFT correspondence.

\section{Thermodynamics of S-AdS BH in Gauss-Bonnet gravity}

In this section, we consider the situation that $R^2$-terms in 
5d action are given by Gauss-Bonnet combination:
\bea
\label{GBaction}
S=\int d^{5}x \sqrt{-g}\left\{ c
\left( \hat{R}^{2}-4\hat{R}_{\mu\nu}\hat{R}^{\mu\nu}
+\hat{R}_{\mu\nu\xi\sigma}\hat{R}^{\mu\nu\xi\sigma} \right)
+{1\over \kappa ^2}\hat{R}-\Lambda \right\} 
\eea
One chooses the coefficients 
$a=c$, $b=-4c$ in the previous action (\ref{vib}) after 
deleting the auxiliary fields. 
The Gauss-Bonnet combination is not topological one if the 
spacetime dimensions are not four. This combination, however, 
appears even in higher dimensions, for example,in the first order 
correction (on string tension) to the low energy effective action in the 
string theory \cite{GBaction}. Furthermore, the Gauss-Bonnet 
combination leads to special structure: if we choose it 
the FRW equations (\ref{R8b}) and (\ref{R9b}) are
\bea
\label{R8bb}
0&=&-{(d-1)k \over 2\kappa^2 r^2} - {(d-1)(d-2) \over 2\kappa^2}
H^2 - \rho \nn
&& - {1 \over 2}(d-1)(d-2)(d-3)(d-4)c H^4 
 - (d-1)(d-3)(d-4) {ckH^2 \over r^2} \nn
&& - {(d-1)(d-3)(d-4) \over 2(d-2)} {ck^2 \over r^4}\ ,\\
\label{R9bb}
0&=&{1 \over \kappa^2}\left\{{(d-3)k \over 2r^4} 
 + {(d-2)H_{,t} \over r^2} 
 + {(d-1)(d-2) \over 2}H^2 \right\} - p \nn
&& + 2(d-2)(d-3)(d-4)c H^2H_{,t} 
+ {1 \over 2}(d-1)(d-2)(d-3)(d-4)c H^4 \nn
&& + {ck \over r^2}\left\{2(d-3)(d-4) H_{,t} + (d-4)(d-3)^2 H^2 \right\} \nn
&& + {(d-1)(d-3)(d-4) \over 2(d-2)}{ck^2 \over r^4}\ .
\eea
Since $H={r_{,t} \over r}$, the derivative terms higher than 
$r_{,t}$ do not appear in (\ref{R8bb}) and the terms higher than 
$r_{,tt}$ do not in (\ref{R9bb}), either. In the sense of 
classical mechanics, if we give the initial values for 
$r$ and $r_{,t}$ which satisfy the ``constraint'' 
(\ref{R8bb}), the time evolutions of these variables can be 
determined by the ``equation of motion'' (\ref{R9bb}). We should 
note that, in general $R^2$-gravity, we need to give the 
initial values for $r$, $r_{,t}$, $r_{,tt}$ and $r_{,ttt}$, 
which satisfy (\ref{R8b}), in order to determine the time 
evolution by (\ref{R9b}). This fact means that there is no 
ghost in the corresponding quantum theory which has  $R^2$ 
terms as the Gauss-Bonnet combination (for very recent 
discussion of such theory see \cite{mgio}), although 
the general higher derivative theory has ghost with negative norm. 

We now treat the theory given by the action (\ref{GBaction}) 
perturbatively  assuming $c\kappa^2$ is small. 
Then the solution for the metric has the same 
form as obtained in $c\ne 0$ case (\ref{dAdS}): 
\bea
\label{dAdS2}
\e^{2\rho}={1\over r^2}\left\{ -\mu +{k\over 2}r^2 + {r^{4}\over l^2}
+{2\mu^2 \epsilon \over r^{4} }\right\}, \quad 
\epsilon = c\kappa^{2}\ .
\eea
We should note, however, the length parameter $l$ is given by
\bea
\label{clc2}
{24 \epsilon \over \kappa^2 l^4 }-{12 \over \kappa^2 l^2}
-\Lambda =0 \ ,
\eea
instead of (\ref{clc}). Then the horizon radius $r_{H}$ and 
the Hawking temperature $T_H$ have the same forms as those 
obtained in (\ref{hrznrds}) and (\ref{ht1}): 
\bea
\label{hrznrds2}
r_{H}&=&r_{0}-{c\mu^{2}\kappa^{2} \over r_{0}^{3}
\left( 2\mu -{k \over 2}r_{0}^{2} \right) }\ ,\quad 
r_{0}^{2}=-{k l^{2} \over 4} + {1\over 2}
\sqrt{{k^2 \over 4}l^{4}+ 4\mu l^{2} }\ , \\
\label{ht12}
T_H &=& {(\e^{2\rho})'|_{r=r_{H}} \over 4\pi} \\
&=& {1 \over 4\pi}\left\{ {4r_{H} \over l^2 }+ {k\over r_{H}}
-{8\epsilon \over r_{H}^{7} } \left({k\over 2}r_{H}^{2}
+ {r_{H}^{4} \over l^2} 
\right)^{2} \right\} \; , \nonumber
\eea
The free energy $F$ looks as follows:
\bea
F &=& {V_{3} \over \kappa^2 l^2}\left[
{l^2 k r_{H}^2 \over 2}+\epsilon \left\{
{14 r_{H}^{4} \over l^{2}} -4r_{H}^{2} k
+{l^{2}k^{2} \over 2 } \right.\right. \nn
&& \left.\left. + {12 l^2 \over r_H^4}
\left( {k \over 2}r_{H}^{2}
+{r_{H}^{4} \over l^{2} } \right)^{2} 
\right\}\right] \ ,
\eea
The entropy ${\cal S}$ and the energy $E$ have the following form:
\bea
\label{entGB}
{\cal S} &=& {4 V_{3}\pi r_{H}^{3} \over \kappa^{2} }
\left[ 1 -  {12\epsilon \over l^{2}} \left\{ 1 
+ {c\kappa^2 \over 8l^{2}} 
\left( 4 + {kl^2 \over r_H^2}\right)\left(2 
+ {k l^2 \over r_H^2} \right)\right.\right. \nn
&&\left.\left. \times \left( 1-{ kl^{2} \over 4r_{H}^{2} }
\right)^{-1} \right\}\right] \ ,\\
\label{enerGB}
E&=&{3 V_{3} \over \kappa^{2} }\left[ \mu 
 -  {\epsilon \over l^{2}} \left\{ 12 \mu 
+ \left(k^2 l^2 + 16 \mu\right)\left(k l^2 
+ \sqrt{k^2 l^4 + 16 l^2\mu}\right) \right.\right. \nn
&& \left.\left. \times\left(-2 k l^2 + \sqrt{k^2 l^4 
+ 16 l^2\mu}\right)^{-1}\right\}\right]\ .
\eea
Now we again rewrite the metric (\ref{dAdS}) of Schwarzschild-anti 
de Sitter space with correction in the FRW form:
\bea
\label{kappaGB}
&& \kappa_4^2={2 \kappa^2 \over l}\ ,\quad 
{1 \over \tilde \kappa^2}\equiv {1 \over \kappa^2}
 - {12c \over l^2}\ ,\\
\label{F1GB}
&& H^2 = - {k \over 2r^2} + {\kappa_4^2 \over 6}
{\tilde E \over V}\ ,\quad
{\tilde E}={6 V_3 \over \kappa_4^2 r}
\left(\mu - {2\epsilon\mu^2 \over r^4}\right)\ ,\\
\label{2FR1b}
&& \dot H = - {\kappa_4^2 \over 4} \left({\tilde E \over V} 
+ p\right) + {k \over 2r^2}\ ,\quad 
p={2 \over r^4 \kappa_4^2}\left(\mu
 - {10\epsilon\mu^2 \over r^4}\right)\ .
\eea
These equations are identical with the previous 
ones (\ref{F1}) and (\ref{2FR1}) with $a=b=0$ and $c\neq 0$ 
except the length parameter is given by (\ref{clc2}). 
The obtained energy $\tilde E$ in (\ref{F1GB}) is not identical 
with $E$ in (\ref{enerGB}), in general, even if we take account 
of the factor ${l \over r}$ coming from the ratio of the time 
scale in 5d bulk space time and that on the brane:
\bea
\tilde{E}\neq {l \over r}E \ .
\eea
One can find the two energies become equal to each other 
when the brane exists at $r=r_{\rm br}$, which is given by 
(\ref{bbrs1}). 
Similarly, we can also obtain four dimensional entropy 
$\tilde{\cal S}$ by using generalized Cardy formula of 
two dimensional CFT (\ref{CV1}) and (\ref{CV2}). 
The obtained entropy ${\cal S}$ has the same form of 
that in $a=b=0$ and $c \ne 0$ case, and especially, 
when the brane crosses the horizon, we have 
\bea
\label{entCVGB}
\tilde{\cal S}={4\pi (d-2) V \over l\kappa_d^2}
={2\pi (d-2) r_H^{d-1} V_{d-1} \over \tilde \kappa^2}\ .
\eea
By putting $d=4$, we find that the difference between 
${\cal S}$ in (\ref{entGB}), obtained from the bulk black hole, 
and $\tilde{\cal S}$ in (\ref{entCVGB}) is again given by 
(\ref{CV5}). The difference between $\tilde{\cal S}$ 
and ${\cal S}$ is again caused by the breaking of the 
conformal symmetry of QFT dual.

\section{Asymmetrically warped spacetimes and gravitational Lorentz 
violation in $R^2$-gravity}

In this section we will be interested in so-called asymmetrically
warped spacetimes for our theory.
In \cite{CLV} where such backgrounds were considered, it has 
been shown that the apparent violation of the Lorentz invariance 
can occur. They have considered the 
brane in the bulk space which is the charged black hole, 
whose metric is given by
\bea
\label{QSAdS}
ds^2&=&-\e^{2\rho}dt^2 + \e^{-2\rho}dr^2 
+ r^2\sum_{i,j}^3 g_{ij}dx^i dx^j\ ,\nn
\e^{2\rho}&=&{1\over r^2}\left\{ -\mu 
+{k\over 2}r^2 + {r^{4}\over l^2}
+{Q^2 \over r^2 }\right\}\ .
\eea
Here $Q$ is the charge of the black hole. In this section, 
we consider similar black hole background with correction 
coming from the squared Riemann tensor term ($c\neq 0$ case). 
We show that the correction gives the origin for  $Q^2$-like 
term in (\ref{QSAdS}) and the apparent Lorentz violation 
may occur. 

If the photon is confined on the brane, the photon can 
propagate on the brane. If there is a black hole in the 
bulk, the geodesic line does not exist, in general, on the 
brane. Then the massless mode or particle 
propagating in the bulk, such as 
graviton, can propagate faster than the light on the brane. 
This causes the apparent violation of the Lorentz invariance 
\cite{CLV}. 

We now consider the black hole background with the 
correction from $R^2$ (the square of the Riemann tensor) term:
\bea
\label{SAdSwithc}
ds^2&=&\hat G_{\mu\nu}dx^\mu dx^\nu \nn
&=&-\e^{2\rho}dt^2 + \e^{-2\rho}dr^2 
+ r^2\sum_{i,j}^3 g_{ij}dx^i dx^j\ ,\nn
\e^{2\rho}&=&{1\over r^2}\left\{ -\mu 
+{k\over 2}r^2 + {r^{4}\over l^2}
+{2\mu^2 \epsilon \over r^{4} }\right\}, \quad 
\epsilon = c\kappa^{2}\ .
\eea
For static brane solution ($H=0$), the FRW-type 
equation (\ref{F1}) has the following form:
\be
\label{stBR}
0=-{k \over 2r^2} + {\mu \over r^4}
 - {2\epsilon \mu^2 \over r^8}\ .
\ee
For simplicity, the case $k=0$ is considered in this section. 
Then the horizons defined by $\e^{2\rho}=0$ are given by 
\bea
\label{k0hrzn}
r_{H\pm}^4&=&{\mu l^2 \pm \sqrt{\mu^2 l^4 - 8\mu^2 l^2 \epsilon}
\over 2} \nn
&=& \mu l^2\left(1 - {2 \epsilon \over l^2}\right)\ , \quad 
2\mu\epsilon\left(1+{2\epsilon \over l^2}\right)\ .
\eea
Note that there appear two horizons due to $R^2$-term. 
The larger (outer) one $r=r_{H+}$ corresponds to the solution in 
(\ref{hrznrds}). 
In case of $k=0$, the solution $r_0$ of (\ref{stBR}) is given by 
\be
\label{stsl}
r_0^4 = 2 \epsilon \mu\ .
\ee
The brane can exist due to the squared Riemann tensor term.  
Comparing Eq.(\ref{k0hrzn}) and Eq.(\ref{stsl}), we find that the 
brane lies inside the inner horizon. 

Let us assume the brane is static and exists at $r=r_0$. 
Then the velocity $c_{\rm phtn}$ of the photon propagating 
on the brane is given by
\be
\label{cp1}
c_{\rm phtn}(r_0)={\e^{\rho(r_0)} \over r_0}\ ,
\ee
which depends on $r_0$. On the other hand, the geodesic of the 
particle propagating the bulk is given by the Euler-Lagrange 
equation derived from the Lagrangian $L$
\be
\label{cp2}
L={1 \over 2}\left\{-\e^{2\rho(r)}{\dot t}^2 
+ \e^{-2\rho(r)}{\dot r}^2 + r^2 \sum_{i,j}^3 g_{ij}{\dot x}^i 
{\dot x}^j\right\} \ .
\ee
Here $\dot{\ }$ expresses the derivative with respect to the 
proper time $\tau$. If there is non-trivial solution of the 
geodesic, the particle can propagate faster than the 
photon on the brane. 
The Lagrangian has two kinds of the integrals, which 
correspond to the energy $E$ and the momentum $p_i$ of the particle: 
\be
\label{cp3}
-E={\partial L \over \partial \dot t}=-\e^{2\rho(r)}\dot t\ ,
\quad p_i={\partial L \over \partial {\dot x}^i}
=r^2 g_{ij}{\dot x}_j\ .
\ee
For simplicity, one considers the case that the brane is flat: 
$g_{ij}=\delta_{ij}$ ($k=0$). If the particle is massless, we have 
$\left({ds \over d\tau}\right)^2=0$ on the geodesic and we 
find 
\be
\label{cp4}
0=-E^2 \e^{-2\rho(r)} + {\dot r}^2 \e^{-2\rho(r)} + 
{p^2 \over r^2}\ ,
\ee
which can be compared with the classical system of the particle 
with unit mass in the potential $V(r)$:
\bea
\label{cp5}
&& {1 \over 2}{\dot r}^2 + V(r)=\hat E \ ,\nn
&& V(r)={p^2 \e^{2\rho(r)} \over 2r^2}
= {p^2 \over 2}\left\{ - {\mu \over r^4} + {1 \over l^2}
+{2\mu^2 \epsilon \over r^8 }\right\} \ , \nn
&& \hat E={1 \over 2}E^2 \ .
\eea
If $\epsilon=0$ or $\epsilon<0$, the potential is 
monotonically increasing 
function of $r$ and unbounded when $r\rightarrow 0$. Therefore 
if initially $\dot r\leq 0$ or the brane is the boundary of the 
bulk and there is no region with $r>r_0$, by the analogy with 
classical mechanics, the particle radiated from the brane 
cannot return to the brane but falls into the black hole 
singularity. The situation does not change if we consider the 
$R^2$-gravity without the Riemann tensor square term, that is, 
the gravity including the squares of the scalar curvature 
and Ricci tensor, where HD correction is included into the 
redefinition of the radius parameter $l$.

On the other hand, if $\epsilon >0$, the potential 
increases when $r$ is small. Since we are treating $\epsilon$ as 
a parameter of the perturbation, we cannot say any definite thing 
when $r$ is small, where the correction becomes large, but the 
potential could be bounded and there would be non-trivial geodesic 
line, along which the particle can return to the brane, as 
in the case of the charged black hole \cite{CLV}.
Hence, HD gravity may lead to brane-worlds with apparent violations 
of Lorentz invariance as it occurs also in Einstein-Maxwell 
gravity \cite{CLV}.

\section{Brane solutions in de Sitter space}

It is right time now to look to other type of bulk space,
i.e. to de Sitter space. The motion of domain walls in 
de Sitter bulk has been recently discussed in ref.\cite{strings}. 
Moreover, there appeared recently attempt to formulate 
dS/CFT correspondence \cite{strominger,dscft}. There were also 
earlier proposals on dS/CFT duality \cite{desitter} and 
thermodynamics of de Sitter space was always under attention 
(for recent discussion, see \cite{thermodynamics}).
Unfortunately, so far it did not appear any explicit example 
of dual CFT for de Sitter space (except not completely physical 
example by Hull \cite{desitter}). Nevertheless, the attempts 
to find such example continue. 

In \cite{dsBrane}, the quantum creation 
of four dimensional de Sitter brane universe in five dimensional 
de Sitter spacetime as in  scenario of so-called Brane New 
World \cite{NOZ,HHR0} is investigated. For $R^2$-gravity, 
if one of the solutions for $l^2$ in Eq.(\ref{ll}) is negative:
\be
\label{Bds1}
l_{\rm dS}^2= -l^2 >0\ ,
\ee
de Sitter space is  exact solution. Moreover, even 
if the cosmological constant $\Lambda$ is negative, there can 
be a de Sitter space solution due to $R^2$ terms. 
By the Wick rotation, 
5d de Sitter space becomes 5d sphere, whose metric is given by
\be
\label{dSi}
ds^2_{{\rm S}_5}=dz^2 + l_{\rm dS}^2 \sin^2 
{z \over l_{\rm dS}}d\Omega^2_4\ .
\ee
Here $d\Omega^2_4$ describes the metric of ${\rm S}_4$ 
with unit radius. The coordinate $z$ is defined in 
$0\leq z \leq l_{\rm dS}\pi$. One also assumes the brane lies 
at $z=z_0$ and the bulk space is given by gluing two regions 
given by $0\leq y < y_0$. Identifying 
\be
\label{Bds2}
A=\ln \sin {z \over l_{\rm dS}}\ 
\ee
and using (\ref{x}), we obtain the following equation:
\be
\label{Bds3}
0={1 \over R}\sqrt{-1 + {R^2 \over l_{\rm dS}^2}}
 - {1 \over l_{\rm dS}}\ .
\ee
Here $R\equiv l_{\rm dS} \sin {y_0 \over l_{\rm dS}}$ is 
the radius of the brane. 
In (\ref{Bds3}), the contribution from $R^2$-terms appears through 
$l$ by (\ref{ll}). 
When bulk is AdS, we need the quantum contribution 
from the matter on the brane in order that de Sitter brane 
existed. When  bulk is de Sitter, even in classical case that 
there is no quantum contribution 
from the matter on the brane, Eq.(\ref{Bds3}) has a solution:
\be
\label{Csol}
R^2=R_0^2\equiv 
{l_{\rm dS}^2 \over 2}\ \mbox{or}\ {y_0 \over l_{\rm dS}}
={\pi \over 6}\ ,\quad {5\pi \over 6}\ .
\ee
In Eq.(\ref{Bds3}), the first term 
corresponds to the gravity, which makes the radius $R$ larger. 
On the other hand, the second term corresponds to the tension, 
which makes $R$ smaller. When $R<R_0$,  gravity becomes larger 
than the tension and when $R>R_0$, vice versa. Then both of 
the solutions in (\ref{Csol}) are stable. Although it is not 
clear from (\ref{Bds3}), $R=l_{\rm dS}$ (${y \over l_{\rm dS}}
={\pi \over 2}$) corresponds to the local maximum. 

When one considers brane in de Sitter space, there is no any 
essential difference between  Einstein gravity and 
$R^2$-gravity. Then we now consider the black hole solution 
when $c\neq 0$ and the FRW type equation (\ref{F1}) exists:
\bea
\label{F1b} 
H^2 &=& {2 \over l_{\rm dS}^2} - {1 \over r^2} 
+ {\mu \over r^4}  - {2\epsilon\mu^2 \over r^8}\ .
\eea
The last term in (\ref{F1b}) comes from $R^2$-term. 
One cannot embed the brane with the shape of 
flat plane or hyperboloid in the bulk sphere, then 
it is reasonable to consider  brane with the shape of 
sphere $k=2$. The static solution corresponds to $H=0$. 
When $\epsilon=0$ ($c=0$), the static solution 
\be
\label{BB2}
r^2=r_{0\pm}^2={l_{\rm dS}^2 \pm \sqrt{l^4 - 8\mu 
l_{\rm dS}^2} \over 4}
\ee
if 
\be
\label{BB3}
8\mu \leq l_{\rm dS}^2\ .
\ee
The case of $8\mu=l_{\rm dS}^2$ corresponds to the extremal case. 
When $\epsilon\neq 0$, where  $\epsilon$ is small,  
using the perturbation with respect to $\epsilon$ one gets 
\bea
\label{Bds4}
r^2&=&r_{0\pm}^2 + {2\epsilon \mu^2 \over r_{0\pm}^4 
 - 2\mu r_{0\pm}^2} \\
&=&{l_{\rm dS}^2 \pm \sqrt{l_{\rm dS}^4 - 8\mu l_{\rm dS}^2} 
\over 4} + {16\epsilon \mu^2 \over 2l_{\rm dS}^4
 - 16\mu l_{\rm dS}^2 \pm \left(2l_{\rm dS}^2
 - 8\mu\right)\sqrt{l_{\rm dS}^4 - 8\mu l_{\rm dS}^2}} \ ,
\nonumber
\eea
when $8\mu < l_{\rm dS}^2$ and 
\be
\label{Bds5}
r^2={l^4 \over 4}+\sqrt{2\epsilon}\ ,
\ee
when $8\mu=l_{\rm dS}^2$. In (\ref{Bds4}), when $\mu$ is small 
\be
\label{Bds6}
r^2 - r_{0\pm}^2={4\epsilon \mu^2 \over l^4}\ ,\quad -\epsilon\ .
\ee
Therefore if $\epsilon$ is positive, the correction from 
$R^2$(Riemann tensor)-term makes the radius of the outer (inner) 
horizon larger (smaller). 

We will finish with this explicit example of dS brane in dS bulk. 
Of course, one can go to details and to discuss brane FRW-dynamics
as in section 5, etc.
For our purposes the explicit demonstration of existence of dS 
brane in dS bulk is enough.

\section{Conformal Anomaly from $R^2$-gravity in 
dS/CFT correspondence}

As it was mentioned in previous section, the explicit example 
of dS/CFT correspondence still did not appear. Nevertheless, 
it has been demonstrated in ref.\cite{strominger} and in 
third paper from \cite{dscft} that supposing dS/CFT correspondence 
the central charge (conformal anomaly) of dual CFT may 
be derived from dS gravity.

In \cite{anom} conformal anomaly from $R^2$-gravity in AdS/CFT 
correspondence has been found. In this section, using similar 
method we derive conformal anomaly in dS/CFT correspondence. 
Let us write $l_{\rm dS}$ in (\ref{Bds1}) as $l$, therefore 
instead of (\ref{ll}), $l$ should satisfy the following equation:
\be
\label{llds}
0={80 a \over l^4} + {16 b \over l^4} + {8c \over l^4}
+ {12 \over \kappa^2 l^2}-\Lambda\ .
\ee

In order to calculate the conformal anomaly from bulk dS gravity, 
one considers the fluctuations around de Sitter space, whose 
metric can be expressed as 
\be
\label{dsmtrc}
ds_{\rm dS}^2=-{l^2 \over 4}
\rho^{-2}d\rho^2 + \rho^{-1}\sum_{i=1}^d \left(dx^i\right)^2\ .
\ee
Then analogously to AdS case \cite{HS}, 
we assume the metric has the following form:
\bea
\label{ai}
ds^2&\equiv&\hat G_{\mu\nu}dx^\mu dx^\nu 
= -{l^2 \over 4}\rho^{-2}d\rho^2 + \sum_{i=1}^d
\hat g_{ij}dx^i dx^j \ ,\nn
\hat g_{ij}&=&\rho^{-1}g_{ij}\ .
\eea
Suppose that there is a brane at $\rho=0$. 
Note that there is a redundancy in the expression of 
(\ref{ai}). In fact, if we reparametrize the metric :
\be
\label{ia}
\delta\rho= \delta\sigma\rho\ ,\ \ 
\delta g_{ij}= \delta\sigma g_{ij}\ .
\ee
by a constant parameter $\delta\sigma$, the expression (\ref{ai}) 
is invariant. The transformation (\ref{ia}) is nothing but 
the scale transformation on the brane. 

In the parametrization (\ref{ai}), scalar curvature $\hat R$ 
\bea
\label{aiii}
\hat R&=&{d^2 + d \over l^2}+\rho R 
-{2(d-1) \rho \over l^2}g^{ij}g'_{ij}
- {3\rho^2 \over l^2} g^{ij}g^{kl}g'_{ik}g'_{jl} \nn
&& + {4\rho^2 \over l^2}g^{ij}g''_{ij}
+ {\rho^2 \over l^2}g^{ij}g^{kl}g'_{ij}g'_{kl} 
\eea
Ricci tensor $\hat R_{\mu\nu}$
\bea
\label{aiv}
\hat R_{\rho\rho}&=&-{d \over 4\rho^2} - {1 \over 2}g^{ij}g''_{ij} 
+ {1 \over 4}g^{ik}g^{lj}g'_{kl}g'_{ij}\ , \nn
\hat R_{ij}&=&R_{ij} + {2\rho \over l^2}g''_{ij} 
- {2\rho \over l^2}g^{kl}g'_{ki}g'_{lj} 
+ {\rho \over l^2}g'_{ij}g^{kl}g'_{kl} \nn
&& + {2-d \over l^2}g'_{ij} - {1 \over l^2}g_{ij}g^{kl}g'_{kl} 
+ {d \over l^2\rho}g_{ij} \ ,\nn
\hat R_{i\rho}&=& \hat R_{\rho i} \\
&=&{1 \over 2}g^{jk}g'_{ki,j} - {1 \over 2}g^{kj}g'_{jk,i} 
+ {1 \over 2}g^{jk}_{,j}g'_{ki} 
+ {1 \over 4}g^{kl}g'_{li}g^{jm}g_{jm,k} 
- {1 \over 4}g^{kj}_{,i}g'_{jk} \ ,\nonumber
\eea
Riemann tensor $\hat R_{\mu\nu\rho\sigma}$
\bea
\label{av}
\hat R_{\rho\rho\rho\rho}&=& 0 \ ,\nn
\hat R_{\rho\rho\rho i}&=&R_{\rho\rho i\rho}=R_{\rho i\rho\rho}
=R_{i \rho\rho\rho}=0 \ ,\nn
\hat R_{ij\rho\rho}&=&\hat R_{\rho\rho ij}=0 \ ,\nn
\hat R_{i\rho j\rho}&=&\hat R_{\rho i\rho j}=
-\hat R_{i\rho \rho j}=\hat R_{\rho ij \rho} \nn
&=& - {1 \over 4\rho^3}g_{ij} + {1 \over 4\rho}g^{kl}g'_{ki}g'_{lj}
- {1 \over 2\rho}g''_{ij} \ ,\nn
\hat R_{\rho ijk}&=& - \hat R_{i \rho jk} = \hat R_{jk \rho i}
= - \hat R_{jki \rho} \nn
&=& {1 \over 4\rho }\left\{ 2g'_{ij,k} - 2g'_{ik,j} 
- g^{lm}\left(g_{im,k} + g_{km,i} - g_{ik,m}\right)g'_{lj} 
\right. \nn
&& \left. + g^{lm}\left(g_{im,j} + g_{jm,i} - g_{ij,m}\right)g'_{lk} 
\right\} \ ,\nn
\hat R_{ijkl}&=&{1 \over \rho}R_{ijkl} \\
&& + {1 \over \rho^2 l^2}\left\{\left( g_{jl}-\rho g'_{jl}\right)
\left(g_{ik} - \rho g'_{ik}\right)
-\left( g_{jk}-\rho g'_{jk}\right)
\left(g_{il} - \rho g'_{il}\right)\right\} \ .\nonumber
\eea
may be easily calculated.
Here ``$\ '\ $" expresses the derivative with respect to $\rho$ 
and $R$, $R_{ij}$, $R_{ijkl}$ are scalar curvature, Ricci and 
Riemann tensors, respectively, on $M_d$.

As in the previous papers \cite{HS} on holographic conformal 
anomaly, we expand the metric $g_{ij}$ as a power 
series with respect to $\rho$,
\be
\label{vii}
g_{ij}=g_{(0)ij}+\rho g_{(1)ij}+\rho^2 g_{(2)ij}+\cdots \ .
\ee
Substituting (\ref{vii}) into (\ref{aiii}), (\ref{aiv}) 
and (\ref{av}), one gets
$\rho$:
\bea
\label{bi}
\lefteqn{\sqrt{-\hat G}={l \over 2}\rho^{-{d \over 2}-1}
\sqrt{g_{(0)}}\left\{1 + {\rho \over 2}g_{(0)}^{ij}g_{(1)ij} 
+ \rho^2\left({1 \over 2}g_{(0)}^{ij}g_{(2)ij} \right.\right.} \nn
&& \left.\left. - {1 \over 4}g_{(0)}^{ij}g_{(0)}^{kl}
g_{(1)ik} g_{(1)jl} 
+ {1 \over 8}\left(g_{(0)}^{ij}g_{(1)ij} \right)^2\right)
+ O(\rho^3)\right\} \ ,\nn
\lefteqn{\sqrt{-\hat G}\hat R={l \over 2}\rho^{-{d \over 2}-1}
\sqrt{g_{(0)}}\left\{{d^2 + d \over l^2} \right. 
+ \rho \left(R_{(0)}
 - {-d^2+3d-4 \over 2l^2}g_{(0)}^{ij}g_{(1)ij}\right)}  \nn
&& + \rho^2\left( -g_{(1)ij}R_{(0)}^{ij}
+ {1 \over 2}R_{(0)}g_{(0)}^{ij}g_{(1)ij} \right. 
- {-d^2 + 7d -24 \over 2l^2}g_{(0)}^{ij}g_{(2)ij} \nn
&& - {d^2 - 7d + 20 \over 4l^2}
g_{(0)}^{ij}g_{(0)}^{kl}g_{(1)ik} g_{(1)jl} 
\left.\left. 
- {-d^2 + 7d - 16 \over 8l^2}\left(g_{(0)}^{ij}g_{(1)ij} \right)^2
\right)+ O(\rho^3)\right\} \ ,\nn
\lefteqn{\sqrt{-\hat G}\hat R^2={l \over 2}\rho^{-{d \over 2}-1}
\sqrt{g_{(0)}}\left\{{d^2(d+1)^2 \over l^4} \right. 
+ \rho \left( {2d(d+1) \over l^2}R_{(0)} 
\right. }\nn
&& \left.
+ {d^4 -6d^3 + d^2 + 8d \over 2l^4}g_{(0)}^{ij}g_{(1)ij}\right) 
+ \rho^2\left( R_{(0)}^2
 - {2d(d+1) \over l^2}g_{(1)ij}R_{(0)}^{ij} \right. \nn
&& - {-d^2 + 3d -4 \over l^2}R_{(0)}g_{(0)}^{ij}g_{(1)ij} 
+ {d^4 -14d^3 +33d^2 + 48d \over 2l^4}g_{(0)}^{ij}g_{(2)ij} \nn
&& + {-d^4 + 14d^3 -25d^2 - 40d \over 4l^4}
g_{(0)}^{ij}g_{(0)}^{kl}g_{(1)ik} g_{(1)jl} \nn
&& \left.\left. + {d^4 -14d^3 +49d^2 -32d + 32 \over 8l^4}
\left(g_{(0)}^{ij}g_{(1)ij} \right)^2
\right)+ O(\rho^3)\right\}\ , \nn
\lefteqn{\sqrt{-\hat G}\hat R_{\mu\nu} \hat R^{\mu\nu}
={l \over 2}\rho^{-{d \over 2}-1}
\sqrt{g_{(0)}}\left\{{d^2(d+1) \over l^4} \right. 
+ \rho \left( {2d \over l^2}R_{(0)}
\right.}\nn 
&& \left. + {d^3 -7d^2 + 8d \over 2l^4}g_{(0)}^{ij}g_{(1)ij}\right) 
+ \rho^2\left( R_{(0)}^{ij}R_{(0)ij} 
 - {4d -4 \over l^2}g_{(1)ij}R_{(0)}^{ij} \right. \nn
&& - {-d +2 \over l^2}R_{(0)}g_{(0)}^{ij}g_{(1)ij} 
+ {d^3 -15d^2 + 48d  \over 2l^4}g_{(0)}^{ij}g_{(2)ij} \nn
&& + {-d^3 +19d^2 - 56d +16 \over 4l^4}
g_{(0)}^{ij}g_{(0)}^{kl}g_{(1)ik} g_{(1)jl} \nn
&&  \left.\left. + {d^3 -15d^2 +56d -32 \over 8l^4}
\left(g_{(0)}^{ij}g_{(1)ij} \right)^2
\right)+ O(\rho^3)\right\} \ ,\nn
\lefteqn{\sqrt{-\hat G}\hat R_{\mu\nu\rho\sigma} 
\hat R^{\mu\nu\rho\sigma}
={l \over 2}\rho^{-{d \over 2}-1}
\sqrt{g_{(0)}}\left\{{2d(d+1) \over l^4} \right. 
+ \rho \left( {4 \over l^2}R_{(0)} \right.}\nn 
&& \left. + {d^2 -7d +8 \over 2l^4}g_{(0)}^{ij}g_{(1)ij}\right)  
+ \rho^2\left( R_{(0)}^{ijkl}R_{(0)ijkl} 
- {4 \over l^2}g_{(1)ij}R_{(0)}^{ij}
+ {2 \over l^2}R_{(0)}g_{(0)}^{ij}g_{(1)ij} \right. \nn
&& + {d^2 -15d +48 \over l^4}g_{(0)}^{ij}g_{(2)ij} 
+ {-d^2 +23d -56 \over 2l^4}
g_{(0)}^{ij}g_{(0)}^{kl}g_{(1)ik} g_{(1)jl} \nn
&& \left.\left. + {d^2 -15d + 48 \over 4l^4}
\left(g_{(0)}^{ij}g_{(1)ij} \right)^2
\right) + O(\rho^3)\right\} \ .
\eea
We regard $g_{(0)ij}$ in (\ref{vii}) as independent field on 
the brane or boundary, which we denote by $M_d$. 
One can solve $g_{(l)ij}$ ($l=1,2,\cdots$) with respect to 
$g_{(0)ij}$  using equations of motion. 
When substituting the expression (\ref{vii}) or (\ref{bi}) 
into the classical action (\ref{vi}), 
the action contains the divergence from $\rho\rightarrow 0$ 
in general situation. 
We regularize the divergence by introducing a 
cutoff parameter $\epsilon$:
\be
\label{aviii}
\int d^{d+1}x\rightarrow \int d^dx\int_\epsilon d\rho \ ,\ \ 
\int_{M_d} d^d x\Bigl(\cdots\Bigr)\rightarrow 
\int d^d x\left.\Bigl(\cdots\Bigr)\right|_{\rho=\epsilon}\ .
\ee
Then the action (\ref{vi}) can be expanded as a power series of 
$\epsilon$:
\bea
\label{viiia}
S&=&S_0(g_{(0)ij})\epsilon^{-{d \over 2}}
+ S_1(g_{(0)ij}, g_{(1)ij})\epsilon^{-{d \over 2}-1} \nn
&& + \cdots + S_{\rm ln} \ln \epsilon 
 - S_{d \over 2} + {\cal O}(\epsilon^{1 \over 2}) \ .
\eea
The term $S_{\rm ln}$ proportional to $\ln\epsilon$ 
appears when $d=$even. 
In (\ref{viiia}), the terms proportional to the inverse power of 
$\epsilon$ in the regularized action are invariant under the scale 
transformation 
\be
\label{viiib}
\delta g_{(0)\mu\nu}=2\delta\sigma g_{(0)\mu\nu}\ ,\ \  
\delta\epsilon=2\delta\sigma\epsilon \ . 
\ee
The invariance is caused by (\ref{ia}). 
The subtraction of these terms proportional to the 
inverse power of $\epsilon$ does not break the invariance.
When $d$ is even, however, there appears the term 
$S_{\rm ln}$ proportional to $\ln\epsilon$.
The subtraction of the term $S_{\rm ln}$ breaks 
the invariance under the transformation (\ref{viiib}). 
The reason is that the variation of the $\ln\epsilon$ term 
under the scale transformation (\ref{viiib}) is finite 
when $\epsilon\rightarrow 0$ since 
$\ln\epsilon \rightarrow \ln\epsilon + \ln(2\delta\sigma)$. 
Therefore the variation should be cancelled by the variation of 
the finite term $S_{d \over 2}$ (which does not depend 
on $\epsilon$) 
\be
\label{vari}
\delta S_{d \over 2}=\ln (2\sigma)S_{\rm ln}
\ee
since the original total action (\ref{vi}) is invariant under the 
scale transformation. 
Since the action $S_{d \over 2}$ can be regarded as the action 
renormalized by the subtraction of the terms which diverge when 
$\epsilon\rightarrow 0$, 
the $\ln\epsilon$ term $S_{\rm ln}$ gives the conformal anomaly $T$ 
of the renormalized theory on the boundary $M_d$: 
\be
\label{xi}
S_{\rm ln}={1 \over 2}
\int d^dx \sqrt{-g_{(0)}}T \ .
\ee
When $d=4$, by substituting the expressions in (\ref{bi}) into the 
action (\ref{vi}) and using the regularization in (\ref{aviii}), 
we find 
\bea
\label{xiii}
S_{\rm ln}&=&-{1 \over 2}\int dx^4\sqrt{-g_{(0)}}\left[
l \left(aR_{(0)}^2 + bR_{(0)}^{ij}R_{(0)ij}
+ c R_{(0)}^{ijkl}R_{(0)ijkl}
\right) \right. \nn
&& + \left( {40a \over l^3} + {8b \over l^3} 
+ {4c \over l^3} + {6 \over l \kappa^2} - {l\Lambda \over 2}\right)
g_{(0)}^{ij}g_{(2)ij} \nn
&& - \left({40a \over l} + {12b \over l} + {12c \over l} 
+ {l \over \kappa^2}\right)g_{(1)ij}R_{(0)}^{ij} \nn
&& - \left(-{8a \over l} - {2b \over l} - {2c \over l} 
 - {l \over 2\kappa^2}\right)R_{(0)}g_{(0)}^{ij}g_{(1)ij} \nn
&& + \left({20a \over l^3} + {8b \over l^3} 
+ {10c \over l^3} - {2 \over l\kappa^2} + {l\Lambda \over 4}\right)
g_{(0)}^{ij}g_{(0)}^{kl}g_{(1)ik}g_{(1)jl} \nn
&& \left. + \left({6a \over l^3} + {2b \over l^3} 
+ {c \over l^3} + {1 \over 2l\kappa^2} - {l\Lambda \over 8}\right)
\left(g_{(0)}^{ij}g_{(1)ij}\right)^2 \right]\ .
\eea
All this reasoning is very similar to the one in AdS/CFT 
correspondence. 

Other terms proportional to $\epsilon^{-1}$ or $\epsilon^{-2}$ 
which diverge when $\epsilon\rightarrow 0$ can be subtracted without 
loss of the general covariance and scale invariance. 
The equation obtained by the variation over $g_{(1)ij}$ is given by
\bea
\label{g1i}
0&=&AR_{(0)}^{ij} + B g_{(0)}^{ij}R_{(0)} 
+ 2C g_{(0)}^{ik}g_{(0)}^{jl}g_{(1)kl}
+ 2D g_{(0)}^{ij}g_{(0)}^{kl}g_{(1)kl} \ ,\nn
A&\equiv&-{40a \over l} - {12b \over l} - {4c \over l} 
- {l \over \kappa^2} \ ,\nn
B&\equiv&{8a \over l} + {2b \over l} + {2c \over l} 
+ {l \over 2\kappa^2} \ ,\nn
C&\equiv& {20a \over l^3} + {8b \over l^3} 
+ {10c \over l^3} - {2 \over l\kappa^2} + {l\Lambda \over 4} 
={40a \over l^3} + {12b \over l^3} 
+ {12c \over l^3} + {1 \over l\kappa^2}\ ,\nn
D&\equiv& {6a \over l^3} + {2b \over l^3} 
+ {c \over l^3} + {1 \over 2l\kappa^2} - {l\Lambda \over 8}
=-{4a \over l^3}  - {1 \over l\kappa^2} \ .
\eea
Here (\ref{llds}) is used in order to rewrite the expressions 
of $C$ and $D$ and to remove the cosmological constant $\Lambda$. 
Multiplying $g_{(1)ij}$ with (\ref{g1i}), we obtain
\be
\label{g1ii}
g_{(0)}^{ij} g_{(1)ij}=-{A+4B \over 2(C+4D)}R_{(0)}\ .
\ee
Substituting (\ref{g1ii}) into (\ref{g1i}), we 
can solve (\ref{g1i}) with respect to $g_{(1)ij}$ as follows:
\be
\label{xiv}
g_{(1)ij}=-{A \over 2C}R_{(0)ij}
+ {AD - BC \over 2C(C+4D)}R_{(0)}g_{(0)ij} \ .
\ee
Substituting (\ref{xiv}) into (\ref{xiii}), 
one finds the following expression for the anomaly $T$
\bea
\label{xv}
T&=&-\left( al + 
{A^2D - 4B^2C - 2ABC \over 4C(C+4D) }\right) R_{(0)}^2 \nn
&& - \left(bl - {A^2 \over 4C}\right)
R_{(0)ij}R_{(0)}^{ij} - cl R_{(0)ijkl}R_{(0)}^{ijkl}\ .
\eea
Substituting (\ref{g1i}) into (\ref{xv}), one gets 
\be
\label{xvaa}
T=-\left({l^3 \over 8\kappa^2}+5al+bl\right)(G-F) 
 - {cl \over 2}(G+F)\ .
\ee
Here  Gauss-Bonnet invariant $G$ and the square of Weyl tensor $F$, 
which are given in (\ref{R12}) appeared. 

We should note that the relative signs in (\ref{xvaa}) 
of the term coming from the Einstein term proportional to 
${1 \over \kappa^2}$ and the terms coming from  $R^2$ terms 
proportional to $a$, $b$ or $c$ are different from the expression 
of the anomaly \cite{anom} in AdS/CFT. The apparent reason is 
that the length parameter $l_{\rm dS}$ in de Sitter space is 
related with that $l_{\rm AdS}$ in AdS by $l_{\rm dS}^2 = - 
l_{\rm AdS}^2$. Of course the difference can be absorbed into the 
redefinition of the parameters $a$, $b$ and $c$. We also note that 
there is an ambiguity of the sign when we relate, in (\ref{viiia}) 
the action in the bulk with Lorentzian signature with that 
in the boundary with Euclidean signature. 

As some speculative example we take ${\cal N}=2$ SCFT theory 
with $n_V$ vector multiplets and $n_H$ hypermultiplets where 
QFT conformal anomaly is given by 
\bea
\label{bng1}
T&=&{1 \over 24\cdot 16\pi^2}\Bigl[ -{1 \over 3}
(11n_V + n_H) R^2 \nn
&& + 12n_V R_{ij}R^{ij} + (n_H-n_V)R_{ijkl}R^{ijkl}\Bigr]\ ,
\eea
and especially, when the gauge group is $Sp(N)$, 
\bea
\label{bng2}
T&=&{1 \over 24\cdot 16\pi^2}\Bigl[ 
-\left(8N^2 + 6N - {1 \over 3}\right) R^2 \nn
&& + (24N^2+12N) R_{ij}R^{ij} + (6N-1)R_{ijkl}R^{ijkl}\Bigr]\ .
\eea
In the framework of AdS/CFT correspondence, 
the ${\cal N}=2$ theory with the gauge group $Sp(N)$ arises as 
the low-energy theory on the world volume on $N$ D3-branes 
sitting inside 8 D7-branes at an O7-plane \cite{Sen}. 
The string theory dual to this theory has been conjectured 
to be type IIB string theory on $AdS_5\times X^5$ where 
$X_5=S^5/Z_2$ \cite{FS}, whose low energy effective action 
is given by  
\be
\label{bng3} 
S=\int_{{\rm AdS}_5} d^5x \sqrt{G}\left\{{N^2 \over 4\pi^2}
\left(R-2\Lambda\right) 
+ {6N \over 24\cdot 16\pi^2}R_{\mu\nu\rho\sigma}
R^{\mu\nu\rho\sigma}\right\}\ .
\ee
In case of AdS/CFT, the conformal anomaly (\ref{bng2}) can be 
reproduced from the action (\ref{bng3}) \cite{anom} to 
the next-to-leading order of ${1 \over N}$. 
For the expression (\ref{xvaa}) derived in the 
framework of dS/CFT, the anomaly (\ref{bng2}) can be 
reproduced to the next-to-leading order of ${1 \over N}$ if we 
choose 
\be
\label{bng4}
{1 \over \kappa^2}={N^2 \over 4\pi^2}\ , \quad
a=b=0\ , \quad c=-{6N \over 24\cdot 16\pi^2}\ .
\ee
Futhermore we also choose
\be
\label{bng5}
\Lambda={12 \over \kappa^2}={12N^2 \over 4\pi^2}\ ,
\ee
to make the length scale $l$ be unity in the leading order 
of ${1 \over N}$. Substituting (\ref{bng4}) 
and (\ref{bng5}) into (\ref{llds}), we find 
\be
\label{bng6}
{1 \over l^2}=
1 - {2c\kappa^2 \over 3} + {\cal O}\left((c\kappa^2)^2\right) \ .
\ee
Then conformal anomaly looks 
\bea
\label{bng7}
T&=&\left(-{1 \over 8\kappa^2}+{c \over 8}\right)(G-F) 
+ {c \over 2}(G+F) + {1 \over \kappa^2}{\cal O}
\left((c\kappa)^2\right) \nn
&=&{N^2 \over 16\pi^2}\left(-{1 \over 3}R_{(0)}^2 
+ R_{(0)ij}R_{(0)}^{ij}\right) \nn
&& + {6N \over 24\cdot 16\pi^2}\left({3 \over 4}R_{(0)}^2 
-{13 \over 4}R_{(0)ij}R_{(0)}^{ij} + R_{(0)ijkl}R_{(0)}^{ijkl}
\right) + {\cal O}(1)\ .
\eea
In the first line in (\ref{bng7}), we substituted (\ref{bng6}) 
and putted $a=b=0$ and in the second line, (\ref{bng4}), 
(\ref{bng5}) and the explicit expression for $G$ and $F$ 
in (\ref{R12}) are substituted. Eq. (\ref{bng7}) exactly 
reproduces the result in (\ref{bng1}) as in \cite{anom}. 
We should note, however, the parameters in (\ref{bng4}) and 
(\ref{bng5}) correspond to the action
\be
\label{bng3b} 
S=\int_{{\rm dS}_5} d^5x \sqrt{G}\left\{{N^2 \over 4\pi^2}
\left(R-2\Lambda\right) 
 - {6N \over 24\cdot 16\pi^2}R_{\mu\nu\rho\sigma}
R^{\mu\nu\rho\sigma}\right\}\ .
\ee
Comparing (\ref{bng3b}) with (\ref{bng3}), one sees the 
sign on front of $R_{\mu\nu\rho\sigma}
R^{\mu\nu\rho\sigma}$ term is different besides 
the sign of the cosmological constant. 

Hence, even in situation where we do not know the string 
theoretical framework for dS/CFT correspondence as in AdS/CFT 
set-up, we constructed  5d de Sitter HD gravity action, which 
might be dual to 4d $Sp(N)$ ${\cal N}=2$ super-Yang-Mills theory. 
This speculation may suggest the strategy for search of 
corresponding string (M-theory) configurations.

\section{Discussion}

In the present paper we constructed number of cosmological 
and BH brane-worlds for general model of HD gravity (mainly in five 
dimensions). The essential feature of the theory under 
consideration is the presence of Riemann tensor square term 
(non-zero $c$ case) and not only of curvature square and Ricci 
tensor square terms as usually occurs in four dimensions. 
Thermodynamics of S-AdS BH which is exact solution of theory 
only when $c=0$ is described in detail (perturbation on $c$ is 
used). The entropy, 
energy and free energy are calculated. It is demonstrated that 
if $c$ is not zero the entropy (energy) is not proportional to 
the area (mass). Moreover, in such a case the entropies found 
by different regularization methods do not coincide (usually, 
in Einstein gravity or HD gravity where $c$ is zero they coincide).

The brane equation of motion (bulk is BH) is presented as 
FRW equation. Using AdS/CFT correspondence in the form 
presented  by Verlinde it is shown that dual QFT is not conformal 
theory when Riemann tensor term presents in the theory. 
As a result the holographic (Hubble) entropy does not coincide 
with BH entropy when $c$ is not zero (usually in AdS/CFT set-up 
they do coincide). The conformal symmetry breaking probably is 
responsible for unusual behaviour of BH entropy. 

Asymmetrically warped spacetime (charged black hole) induced 
by $c$-term effect which plays the role of charge is constructed. 
It is interesting that Lorentz invariance violation for such 
background occurs.
It is also presented cosmological dS brane playing the role of the 
boundary which connects two bulk dS spaces. The radius of such 
dS brane which may be useful for construction of inflationary 
universe is found in terms of parameters of HD gravity. Such 
brane may find the applications in framework of proposed dS/CFT 
correspondence. The holographic conformal 
anomaly from five dimensional dS HD gravity is also evaluated.

In general, the results of our work demonstrate how one can find 
different brane-world solutions and describe their properties for HD 
gravity. Of course, there are many interesting topics left for future 
investigation. For example, how to find the dynamical entropy 
bounds for dS brane in the way similar to refs.\cite{EV,others}? 
This may help in better understanding of dS/CFT correspondence 
(if it exists). Another interesting question is related with 
the effect of parameter $c$ to graviton propagator and the 
corresponding trapping of HD gravity.

\section*{Acknowledgements}

The work by SDO has been supported in part by CONACyT 
(CP, Ref.990356 and grant 28454E), in part by RFBR and in part 
by GCFS Grant E00-3.3-461.
The work by S.O. has been supported in part by the Japan Society 
for the Promotion of Science under the Postdoctoral Research 
Programs. The authors thank Yukawa Institute for Theoretical Physics 
at Kyoto University for kind hospitality. Discussions during 
the YITP workshop YITP-W-01-04  ``Quantum Field Theory 2001'' 
were helpful to complete this work.
S.O would like to thank  R.M. Wald and A. Ishibashi for useful 
discussions on BHs in higher derivative gravity. 

\appendix

\section{Comparison with $c=0$ $R^2$-gravity}

In this Appendix for completeness we summarize the results 
with $c=0$ case in the action (\ref{vib}) based on \cite{SSS}. 

When $c=0$, Schwarzschild-anti de Sitter space is an exact solution:
\bea
\label{SAdSA}
ds^2&=&\hat G_{\mu\nu}dx^\mu dx^\nu \nn
&=&-\e^{2\rho_0}dt^2 + \e^{-2\rho_0}dr^2 
+ r^2\sum_{i,j}^{d-1} g_{ij}dx^i dx^j\ ,\nn
\e^{2\rho_0}&=&{1 \over r^{d-2}}\left(-\mu + {kr^{d-2} \over d-2} 
+ {r^d \over l^2}\right)\ .
\eea
The curvatures have the following form:
\be
\label{cvA}
\hat R=-{d(d+1) \over l^2}\ ,\quad 
\hat R_{\mu\nu}= - {d \over l^2}\hat G_{\mu\nu}\ .
\ee
In (\ref{SAdSA}), $\mu$ is the parameter corresponding to mass 
and the scale parameter $l$ is given by solving the following 
equation:
\be
\label{llA}
0={d^2(d+1)(d-3) a \over l^4} + {d^2(d-3) b \over l^4} \nn
- {d(d-1) \over \kappa^2 l^2}-\Lambda\ .
\ee
We also assume $g_{ij}$ corresponds to the Einstein manifold, 
defined by $r_{ij}=kg_{ij}$, where $r_{ij}$ is the Ricci tensor 
defined by $g_{ij}$ and $k$ is the constant. 
By using the method parallel with section 4, we found 
the following thermodynamical quantities:
\bea
\label{freeA}
F&=& -{V_{3} \over 8}r_{H}^{2} \left( {r_{H}^{2} \over l^{2}}
 - {k \over 2} \right)\left( {8 \over \kappa^2} 
 - {320 a \over l^2} -{64 b \over l^2} \right) \; , \\
\label{entA}
{\cal S }&=&{V_{3}\pi r_H^3 \over 2}
\left( {8 \over \kappa^2}- {320 a \over l^2}
 -{64 b \over l^2} \right)\ ,\\
\label{enerA}
E&=& {3V_{3}\mu \over 8}
\left( {8 \over \kappa^2}- {320 a \over l^2}
 -{64 b \over l^2} \right)\ , 
\eea
which seem to tell that the contribution from 
the $R^2$-terms can be absorbed into the redefinition:
\be
\label{B16A}
{1 \over \tilde\kappa^2}={1 \over \kappa^2} - {40a \over l^2} 
 - {8b \over l^2}\ ,
\ee
although this is not true for $c\neq 0$ case. 

The FRW type equations for $c=0$ case correspond to (\ref{e5}) 
and (\ref{2FR1}):
\bea
\label{F1A}
H^2 &=& - {k \over (d-2)r^2} + {\kappa_d^2 \over (d-1)(d-2)}
{\tilde E \over V}\ ,\nn
\dot H &=& - {\kappa_d^2 \over 2(d-2)} \left({\tilde E \over V} 
+ p\right) + {k \over (d-2)r^2}\ ,\nn
{\tilde E}&=&{(d-1)(d-2) \mu V_{d-1} \over \kappa_d^2 r}\ ,\nn
p&=&{(d-2)\mu \over r^d \kappa_d^2}\ ,\nn
V&=&r^{d-1}V_{d-1}\ .
\eea
Here $\kappa_d$ is $d$ dimensional gravitational coupling, 
which is given by
\be
\label{F2A}
\kappa_d^2={2 \tilde\kappa^2 \over l}\ .
\ee
In case of the Einstein gravity ($a=b=c=0$ in (\ref{vi})), 
the equation corresponding to (\ref{F2A}) has the form 
\be
\label{F2EA}
\kappa_{{\rm (Ein)}d}^2={2 \kappa^2 \over l}\ .
\ee
Then the effects of the higher derivative terms, when $c=0$, 
appear through the redefinition of $\kappa^2$ to $\tilde\kappa^2$. 

When $d=4$,  using (\ref{enerA}) we find
\be
\label{F3A}
\tilde E={l \over r}E\ .
\ee
Here $E$ is given by (\ref{enerA}). The factor ${l \over r}$ 
can be understood as the inverse of the factor for the time 
variables between the brane and the bulk in (\ref{eq14}). 

Using  Cardy-Verlinde formula (\ref{CV1}), one 
has the expression for the entropy $\tilde{\cal S}$ for $c=0$ case 
by using (\ref{CV2}). Especially we can evaluate holographic 
(Hubble) entropy $\tilde{\cal S}$ 
when the brane crosses the horizon $r=r_H$:
\be
\label{CV4A}
\tilde{\cal S}={4\pi (d-2) V \over l\kappa_d^2}
={2\pi (d-2) r_H^{d-1} V_{d-1} \over \tilde\kappa^2}\ .
\ee
For $d=4$, the entropy $\tilde{\cal S}$ is 
identical with the black hole entropy in (\ref{entA})
\be
\label{CV5A}
\tilde{\cal S}={\cal S}\ .
\ee
The relation (\ref{CV5A}) breaks when $c\neq 0$, as shown 
in (\ref{CV5}).

\end{document}